\documentclass[a4paper,11pt]{article}
\usepackage{jheppub}
\usepackage[utf8]{inputenc}
\usepackage{epsf, array, xcolor}
\usepackage{graphicx}
\usepackage{subfigure}
\usepackage{bbold}
\usepackage{amsmath,amssymb,mathtools}
\usepackage{epigraph}
\usepackage[normalem]{ulem}\usepackage{xcolor}

\usepackage[colorlinks=true,linktocpage=true,citecolor=blue,linkcolor=blue,urlcolor=blue]{hyperref}
\usepackage{textcomp}
\usepackage{color}
\usepackage{slashed}
\usepackage{cancel}
\usepackage{comment}
\usepackage{url}
\usepackage{scalerel}
\usepackage{multicol}
\usepackage{multirow}

\usepackage{tikz}
\usepackage{tikz-feynman}
\tikzfeynmanset{compat=1.0.0}
\pgfdeclarelayer{bg}    
\pgfsetlayers{bg,main}  

\makeatletter
\gdef\@fpheader{}
\makeatother

\preprint{UMN-TH-4604/26}

\title{Off-shell continuation from on-shell recursion }

\author[a,b]{Wenqi Ke,}
\author[a]{Zhen Liu,}
\author[a]{Ishmam Mahbub,}

\affiliation[a]{School of Physics and Astronomy, University of Minnesota, Minneapolis, MN, 55455, USA}

\affiliation[b]{    
IFAE, BIST and Dept.~de Física, Universitat Autònoma de Barcelona, \\08193 Bellaterra, Barcelona, Spain}

\emailAdd{wke@umn.edu}
\emailAdd{zliuphys@umn.edu}
\emailAdd{mahbu008@umn.edu}

\abstract{
Starting from on-shell factorization and the polarization sum of an internal massive vector, we provide an off-shell continuation of the transverse and longitudinal contributions defined at arbitrary virtuality. The construction relies on on-shell constructibility and the three-point off-shell \textit{current constraint} $p \cdot J = \mathcal{O}(m)$. For amplitudes with external longitudinal vectors, we develop a method that isolates the mass-singular contributions associated with the longitudinal polarization, namely terms with power-law singularities in the massless limit. In the examples studied, these terms appear as contact contributions in individual factorization channels, and we identify the conditions under which they cancel. We give an independent understanding, based on massive on-shell amplitudes, of the polarization decomposition and diagram-by-diagram high-energy power counting underlying the Feynman-Diagram gauge and related five-component formulations. We further clarify how higher-point current constraints can emerge from on-shell constructibility and the cancellation of mass singularities. We illustrate the mechanism with explicit four-point examples and discuss the constraints imposed by a smooth massless limit. The work provides a systematic bottom-up method for building massive amplitudes free of mass-singular terms and opens a route toward a broader amplitude-based understanding of field decompositions, connecting massive and massless descriptions and on-shell states with off-shell descriptions.
}

\begin{document}
\maketitle
\flushbottom

\section{Introduction}


On-shell methods build scattering amplitudes directly from their analytic
structure, namely locality, unitarity, and little-group covariance, without
referring to a Lagrangian, a gauge choice, or a field
basis~\cite{Britto:2004ap,Britto:2005fq,Bedford:2005yy,Cachazo:2005ca,Badger:2005zh,Badger:2005jv,Ozeren:2006ft,Boels:2007pj,Arkani-Hamed:2008owk,Cohen:2010mi,Elvang:2013cua,Henn:2014yza,Cheung:2015ota,Cheung:2017pzi,Arkani-Hamed:2017jhn,Franken:2019wqr,Ballav:2020ese,  Wu:2021nmq,Travaglini:2022uwo,Ema:2024vww, Ema:2024rss,Gherghetta:2024tob, Ema:2025qgd, Trott:2026ozo}. A central ingredient of this framework is the set of on-shell recursion relations, which construct higher-point amplitudes from lower-point   building blocks~\cite{Britto:2004ap,Britto:2005fq,Elvang:2013cua}. In recent years, the on-shell formalism has been extended to massive particles of arbitrary spin~\cite{Arkani-Hamed:2017jhn}, and correspondingly, the massive recursion relations have been developed~\cite{
Franken:2019wqr,
Ema:2024vww,Ema:2024rss}.
At tree level, the S-matrix of  spontaneously broken Yang-Mills  theory can be derived from 
\textit{minimal} three-point amplitudes (those with the softest high-energy behavior), together with perturbative unitarity, up to scalar four-point
contact terms that are not fixed by factorization~\cite{Ema:2024rss,Jeong:2026dzt} \footnote{Discussions on applying on-shell methods to Standard Model amplitudes can be found
in Refs. \cite{Cohen:2010mi,Bachu:2019ehv, Liu:2022alx, Christensen:2024bdt, Christensen:2024xzs, Ema:2024rss,Christensen:2024nre,Ni:2024yrr,Ni:2025xkg, Croft:2026eio,Ni:2026wiz,Ni:2026mia,Jeong:2026dzt, Trott:2026cjj,Wang:2026qhv}.}. This is the on-shell
counterpart of the tree-unitarity arguments that historically fixed the gauge
couplings, the gauge-boson self-interactions, and the Higgs couplings of the electroweak (EW) theory~\cite{LlewellynSmith:1973yud,Cornwall:1974km,Lee:1977eg}.

Individual channels in massive gauge theories can contain terms proportional to powers of $E/m$, giving rise to mass singularities in the massless limit. These terms originate, for example, from the external longitudinal polarization vectors $\varepsilon_L^\mu = p^\mu/m + \mathcal{O}(m/E)$ or from the massive-vector propagator, whose numerator contains $q^{\mu} q^{\nu}/m^2_I$. The resulting mass-singular contributions cancel in the full amplitude when the interactions and spectrum admit a smooth massless limit. In a gauge-theory formulation, the Slavnov--Taylor (ST) identities organize these cancellations, while the Goldstone-boson equivalence theorem relates the leading high-energy longitudinal amplitudes to Goldstone amplitudes~\cite{Chanowitz:1985hj,Gounaris:1986cr,Bagger:1989fc,He:1992nga}. Nevertheless, the large terms in individual channels obscure the high-energy power counting and can pose problems in the quasi-collinear limit, where resonant diagrams dominate. In such cases, care must be taken to retain the information needed for gauge cancellations in the full amplitude~\cite{Nardi:2024tce,Dittmaier:2025htf}. Summing large terms of order $(E/m)^{n}$ to obtain a much smaller result can also be numerically demanding.

Here mass singularities denote power-law singularities in the massless limit: negative powers of $\tau$ in the simultaneous limit $m_i=\tau\bar m_i\to0$, at fixed mass ratios and dimensionless couplings, with dimensionful couplings scaled consistently. We consider tree-level amplitudes at generic hard kinematics; soft and collinear singularities and loop infrared logarithms are outside this definition. We call individual terms or contributions mass-singular even when they cancel in the full amplitude. The full amplitude diverges in the massless limit only if uncanceled singularities remain. Unless stated otherwise, the massless limit refers to this prescription.


These difficulties arise from how we decompose the amplitude. 
Refs.~\cite{Chanowitz:1985hj,Wulzer:2013mza,Chen:2016wkt,Chen:2019dkx,Cuomo:2019siu} reorganize the full amplitude
through a five-component representation of the weak bosons, in which the first
four components form a Lorentz four-vector carrying the transverse polarizations
and the longitudinal one with its scalar part subtracted,
$\tilde{\epsilon}^{\mu}_L(p)=\epsilon^{\mu}_L(p)-p^{\mu}/m$, and the fifth
component is the associated Goldstone boson.
The same decomposition underlies the
Feynman-Diagram (FD) gauge of Ref.~\cite{Chen:2022gxv}, where the $5\times5$
off-shell propagator is derived by applying the ST identities to the two
subamplitudes attached to each internal weak-boson line after rewriting the unitary-gauge
propagator.\footnote{Ref.~\cite{Chen:2022gxv} further fixes a light-like
reference vector for the internal decomposition,
$n^{\mu}=\big(\mathrm{sgn}(q^{0}),\,-\vec{q}/|\vec{q}|\big)$, with $q^{\mu}$ the
off-shell internal momentum.} In each case the individual diagrams carry both
massive vector and Goldstone contributions, and none grows with energy faster
than the full amplitude, and the high-energy power counting is clear
diagram by diagram.

In this paper, we provide an on-shell study of the longitudinal degree of freedom of a massive gauge boson using recursion relations, without invoking a Lagrangian at any stage. We first analyze the internal massive-vector line, for which the transverse and longitudinal contributions can be separated on a factorization channel.  We show that at the massive-vector pole $\hat{q}^2 =m_I^2$, the sum
over physical polarizations $ 
\sum_{\lambda}\varepsilon^\mu_\lambda (\hat{q})\varepsilon^{\nu}_\lambda(-\hat{q})
$ of the factorized amplitude can always be rewritten in terms
of objects built from the unshifted momentum $q^{\mu}$.\footnote{Throughout we use hatted notation to denote complex shifted momentum.} To derive the off-shell decomposition, the longitudinal mode requires the
scalar part to be subtracted, and the remaining $q_{\mu}$ terms are removed by
an off-shell \emph{current constraint} on the adjacent subamplitude. The result
defines a decomposition of the internal line into transverse and longitudinal contributions at arbitrary virtuality, with the correct on-shell residue. At the level of four-point amplitudes, we show that the construction relies only on factorization, on-shell constructibility, and the current constraint that follows from the minimality of the three-point amplitudes.

We further explain the diagram-by-diagram power counting of Refs.~\cite{Cuomo:2019siu,Chen:2022gxv} for external longitudinal states. Rewriting the longitudinal three-point amplitudes as in Sec.~\ref{sec:3pt}, we obtain an algorithm that isolates mass-singular contributions in each factorization channel. In the $f\bar f VV$ example, these contributions appear as contact terms. Their cancellation occurs across channels and can require additional Higgs exchange. We discuss the connection to Ward identities and the emergence of a four-point current constraint. This mechanism suggests a bottom-up criterion for constructing higher-point amplitudes with a smooth massless limit; extending the explicit construction to higher multiplicity requires the corresponding current constraints.

Throughout we work at tree level and mainly employ the All-Line Transverse (ALT)
shift of Refs.~\cite{Ema:2024vww, Ema:2024rss, Ema:2025qgd}, keeping the discussion valid for an arbitrary shift
wherever the argument does not depend on the specific deformation. The
construction applies to spontaneously broken gauge theories, with the
electroweak sector of the SM as a special case.
The paper is organized as follows. Section~\ref{sec:2} reviews the
polarization decomposition and the three-point amplitudes used in our
construction. Section~\ref{sec:off-shell} develops the off-shell continuation
of internal massive-vector contributions and illustrates it with the
$ffff$, $VVff$, and $VVVV$ amplitudes.
Section~\ref{sec:ext_long} studies mass singularities associated with external
longitudinal vectors, their cancellation in the $VVff$ amplitude, and the
extension of current constraints to higher points. We conclude in
Sec.~\ref{sec:conclusion}. Appendices~\ref{app:conventions},
\ref{app:alt}, and \ref{app:null-decomposition} collect our spinor-helicity
conventions, details of the ALT shift, and the null decomposition in a
factorization channel, respectively.

\subsection*{Summary of results}
\textbf{Off-shell continuation:}
In Sec. \ref{sec:off-shell}, we start from on-shell factorization through an internal massive vector,
\begin{equation}
 \hat A(z)\underset{z\to z_I}{\sim}
 \frac{\hat J_{L\mu}(z_I)
 \bigl[\hat\Pi_{I,T}^{\mu\nu}(z_I)
      +\hat\Pi_{I,L}^{\mu\nu}(z_I)\bigr]
 \hat J_{R\nu}(z_I)}
 {\hat p_I^2(z)-m_I^2}.
\end{equation}
The vector expressions for the polarization tensors admit polynomial continuations,
\begin{align}
 \hat\Pi_{I,T}^{\mu\nu}(z)
 &=\sum_{\rho=\pm}\hat\epsilon_\rho^\mu(z)
   \hat\epsilon_{\bar\rho}^\nu(z)
 =\Pi_{I,T,0}^{\mu\nu}+O(z),\\
 \hat\Pi_{I,L}^{\mu\nu}(z)
 &=\hat\epsilon_L^\mu(z)\hat\epsilon_L^\nu(z)
 =\Pi_{I,L,0}^{\mu\nu}+O(z).
\end{align}
For the amplitudes studied, the current constraints and the sum over factorization roots allow the pole contributions to be expressed in terms of the following unshifted polarization tensors:
\begin{align}
 \Pi_{I,T,0}^{\mu\nu}
 &=-g^{\mu\nu}
   +\frac{p_I^\mu n_I^\nu+p_I^\nu n_I^\mu}
         {p_I\cdot n_I}
   -\frac{m_I^2 n_I^\mu n_I^\nu}{(p_I\cdot n_I)^2},\\
 \Pi_{I,L,0}^{\mu\nu}
 &=\frac{(p_I-2\eta_I)^\mu(p_I-2\eta_I)^\nu}{m_I^2},
 \qquad
 p_I=\lambda_I+\eta_I,\qquad
 \eta_I^\mu=\frac{m_I^2n_I^\mu}{2p_I\cdot n_I}.
\end{align}
The choice of $n_I^{\mu}$ is discussed in Eq. \eqref
{eq:rn1}. Using the current constraints
$p_I\cdot J_L=m_I A_{\phi,L}$ and
$-p_I\cdot J_R=m_I A_{\phi,R}$, which follow from the input three-point amplitude, 
the longitudinal contribution becomes
\begin{equation}
 J_{L\mu}\Pi_{I,L,0}^{\mu\nu}J_{R\nu}
 =
 \left(A_{\phi,L}-\frac{2\eta_I\cdot J_L}{m_I}\right)
 \left(-A_{\phi,R}-\frac{2\eta_I\cdot J_R}{m_I}\right).
\end{equation}
For the constructible four-point amplitudes studied here, these tensors organize the off-shell exchange together with the contact contributions fixed by recursion. The construction uses the off-shell current constraints of the input three-point amplitudes.

\vspace{2mm}

\noindent\textbf{Isolation of external longitudinal mass singularities:} In section \ref{sec:ext_long}, we show how the mass singularities carried by the longitudinal polarization vectors are isolated channel by channel and cancel in the full amplitude in theories with no mass singularities. For this purpose, we decompose the longitudinal polarization as  $\hat \epsilon_{L}^\mu (\hat p_i)=(\hat p_i^\mu-2 \hat \eta_i^\mu)/m_i$, where $\eta_i^\mu$ scales as $\mathcal{O}(m_i^2)$, and use recursion relations starting with the decomposed three-point amplitudes in
Eqs.~\eqref{eq:LVV}--\eqref{eq:LffV-expanded}. Recall that the ALT shift for a  longitudinal external leg is defined as
\begin{align}
\hat\eta_i^\mu = \eta_i^\mu+\tfrac{z}{2}r_i^\mu , \qquad
\hat p_i^\mu = p_i^\mu+zr_i^\mu , \qquad
 \hat\epsilon_{L}^\mu (\hat p_i)= \epsilon_{L}^\mu (p_i),
\end{align}
Rewriting  $\hat \epsilon_{L}^\mu (\hat p_i)=(\hat p_i^\mu-2 \hat \eta_i^\mu)/m_i$, we can see that each of the two terms individually is deformed by the shift,  but the longitudinal polarization vector itself is left invariant. At the level of the three-point amplitude, the contraction with $\hat p_i^{\mu}/m_i$ is well defined in the massless limit as 
 a consequence of choosing minimally coupled three-point amplitudes.
The unshifted $\eta_i^\mu$ scales as $\mathcal{O}(m_i^2)$, and the part of the amplitude corresponding to it will be well defined in the massless limit for each channel. The shifted $\hat\eta_i^\mu$ contains
$zr_i^\mu/2$, which at the factorization pole scales as $\mathcal{O}(m_i^0)$. As we show, the potentially mass-singular terms in higher-point amplitudes will always originate from the shift-dependent parts of
$\hat{\eta}_i^\mu$ and will contribute as contact terms. This observation will allow us to systematically isolate terms that diverge in the massless limit for each factorization channel. 
\section{Theoretical framework}
\label{sec:2}
\subsection{Polarization decomposition}
\label{subsec:long_pol_decomp}
An on-shell massive momentum $p^\mu$ (with $p^2 =m^2$) can be written as a sum of two null vectors:
\begin{align}
      p^\mu \equiv\lambda^\mu+\frac{m^2}{2 \lambda \cdot n}n^\mu,
    \qquad
    \lambda^2=n^2=0,
    \label{eq:q_lambda_n}
\end{align}
where  $\lambda \cdot n \neq 0$.
Denoting $\eta^{\mu}= \frac{m^2}{2 \lambda \cdot n} n^{\mu}$, we can also express the momentum in the spinor notation:
\begin{equation}
    p_{a\dot a}= p_{\mu}(\sigma^{\mu})_{a\dot a}
    =
    |\lambda\rangle[\lambda|
    +
    |\eta\rangle[\eta| ,
    \qquad
 2  \lambda \cdot  \eta =   \langle \lambda\eta\rangle[\eta\lambda]
    = m^2 .
\end{equation}
Details of the spinor-helicity formalism and the conventions adopted throughout this work are provided in Appendix \ref{app:conventions}. The polarization vectors are:
\begin{align}
    \epsilon_+^{\mu}(p)
    &=
    \frac{\langle \eta|{\sigma}^{\mu}|\lambda]}
         {\sqrt{2}\,\langle \eta\lambda\rangle}, \quad
    \epsilon_-^{\mu}(p)
    =
    \frac{\langle \lambda|{\sigma}^{\mu}|\eta]}
         {\sqrt{2}\,[\lambda\eta]} , \quad \epsilon_L^\mu(p)
    =
    \frac{\lambda^\mu- \eta^\mu}{m}.
    \label{eq: polarization}
\end{align}
 Note that they  are defined up to the choice of the reference vector $n^{\mu}$. 
 These polarization vectors obey
\begin{equation}
    p\cdot \epsilon_\rho=0,
    \qquad
    \epsilon_{\rho}\cdot \epsilon_{\rho'}^*=-\delta_{\rho \rho'},
    \qquad \text{with}\quad 
    \rho, \rho' = {\pm,L}
\end{equation}
In the massless limit $m \rightarrow 0$, taken at fixed $\lambda \cdot n$, both $\lambda^{\mu}$ and $n^{\mu}$ remain finite. The decomposition reduces to $p^\mu\to\lambda^\mu$, i.e., $\lambda^\mu$ becomes the massless momentum itself and $n^\mu$ serves as the reference null vector for the transverse polarizations. The longitudinal polarization behaves as \(\epsilon_L^\mu\sim \lambda^\mu/m\) and diverges as \(m\to0\). We introduce the helicity projectors:
\begin{equation}
    \Pi_{\rho}^{\mu\nu}(p;n)
    \equiv
    \epsilon_\rho^\mu(p)\epsilon_\rho^{*\nu}(p),
    \qquad
    \rho=\pm,L .
\end{equation}
When contracted with an amplitude or a current, these projectors isolate the component associated with the given polarization. More precisely, one finds
\begin{equation}
    \Pi_\rho{}^\mu{}_\alpha\,\Pi_{\rho'}^{\alpha\nu}
    =\epsilon_\rho^\mu\big(\epsilon_\rho^*\cdot\epsilon_{\rho'}\big)\epsilon_{\rho'}^{*\nu}
    =-\delta_{\rho\rho'}\,\Pi_\rho^{\mu\nu},
\end{equation}
Thus, each tensor maps an arbitrary four-vector onto the one-dimensional subspace spanned by $\epsilon_\rho^\mu$. Using $\lambda^\mu=p^\mu-m^2n^\mu/(2p\cdot n)$ and
$\lambda\cdot n=p\cdot n$, we express the tensors entirely in terms of
$p^\mu$ and $n^\mu$. The antisymmetric term is unchanged because
$\varepsilon^{\mu\nu\rho\sigma}n_\rho n_\sigma=0$. For each helicity:
\begin{equation}
    \begin{aligned}
        &\Pi_+^{\mu\nu}(p;n)
=
\frac{1}{2}
\left(
-g^{\mu\nu}
+
\frac{p^\mu n^\nu+p^\nu n^\mu}
     {p\cdot n}
-\frac{m^2 n^\mu n^\nu}{(p\cdot n)^2}
\right)
+
\frac{i}{2\,p\cdot n}\,
\varepsilon^{\mu\nu\rho\sigma}
p_\rho n_\sigma\\&
\Pi_-^{\mu\nu}(p;n)
=
\frac{1}{2}
\left(
-g^{\mu\nu}
+
\frac{p^\mu n^\nu+p^\nu n^\mu}
     {p\cdot n}
-\frac{m^2 n^\mu n^\nu}{(p\cdot n)^2}
\right)
-
\frac{i}{2\,p\cdot n}\,
\varepsilon^{\mu\nu\rho\sigma}
p_\rho n_\sigma\\&\Pi_L^{\mu\nu}(p;n)
=
\frac{p^\mu p^\nu}{m^2}
-
\frac{p^\mu n^\nu+p^\nu n^\mu}
     {p\cdot n}+\frac{m^2 n^{\mu} n^{\nu}}{(p\cdot n)^2}
    \end{aligned}
\label{projectors}\end{equation}
The two transverse projectors are related by complex conjugation, whereas the longitudinal projector is real and symmetric, for real momenta.
The transverse polarization sum is
\begin{equation}
 \Pi_T^{\mu\nu}(p;n)\equiv\Pi_+^{\mu\nu}+\Pi_-^{\mu\nu}
 =-g^{\mu\nu}
 +\frac{p^\mu n^\nu+p^\nu n^\mu}{p\cdot n}
 -\frac{m^2n^\mu n^\nu}{(p\cdot n)^2}.
\end{equation}
Adding the longitudinal contribution gives the usual
massive spin-1 completeness relation:
\begin{align}
   \Pi_T^{\mu\nu}+  \Pi_L^{\mu\nu}
=
-g^{\mu\nu}
+
\frac{p^\mu p^\nu}{m^2}.
\end{align}

\subsection{Three-point amplitudes}
\label{sec:3pt}
In this section, we briefly review the three-point amplitudes used throughout the paper and set up our notation. Readers already familiar with this material may skip ahead to the computation in the next section.

All higher-point amplitudes relevant for this work are constructed from on-shell three-point amplitudes using recursion. We use the following minimal triple-vector and vectorlike fermion--vector amplitudes~\cite{Arkani-Hamed:2017jhn}.\footnote{Throughout this discussion, three-point amplitudes are understood as analytic functions of complexified on-shell momenta satisfying $p_1+p_2+p_3=0$ and $p_i^2=m_i^2$. Specifically, the massless limit considered here is a formal limit in complex momentum space.}\footnote{A general chiral fermion--vector vertex has the form $ A_{f_i f_j V^c}^{(\rho_1\rho_2\rho_3)} = g_{ f f V}^{ijc}\, (T^c)_{ji}\, \bar v_2^{(\rho_2)}(p_2) \slashed{\epsilon}_3^{\, \, (\rho_3)} \left(g_L P_L+g_R P_R\right) u_1^{(\rho_1)}(p_1) $. Here we take $g_L=g_R$ and absorb the common factor into $g_{ffV}^{ijc}$. For unequal chiral couplings, the mass-dependent current constraint must be modified accordingly.}
\begin{align}
	A_{V^aV^bV^c}^{(\rho_1\rho_2\rho_3)}
	&= 
	\vcenter{\hbox{\scalebox{0.70}{
	\begin{tikzpicture}[baseline=(v.base)]
	\begin{feynman}[horizontal=p1 to v]
		\vertex (p1) [label=\({\scriptstyle V_1^a}\)];
		\vertex [below right = of p1, blob, fill=gray] (v) {};
		\vertex [below left = of v,label=270:\({\scriptstyle V_2^b}\)] (p2);
		\vertex [right = of v, label=\({\scriptstyle V_3^c}\)] (p3);
		\diagram*{
			(p1) -- [photon,momentum=\({\scriptstyle p_1}\)] (v)
			     -- [photon,rmomentum'=\({\scriptstyle p_2}\)] (p2),
			(v) -- [photon,rmomentum'=\({\scriptstyle p_3}\)] (p3),
		};
	\end{feynman}
	\end{tikzpicture}
	}}}
	=g f^{abc}
	\bigg[
	(\, \epsilon_1^{(\rho_1)}\cdot\epsilon_2^{(\rho_2)}) \,(p_1-p_2)\cdot\epsilon_3^{(\rho_3)}
	+\mathrm{cycl.}
	\bigg] ,
	\label{eq:VVV}
	\\[1em]
	A_{ f_i f_j V^c}^{(\rho_1 \rho_2 \rho_3)}
	&=
	\vcenter{\hbox{\scalebox{0.70}{
	\begin{tikzpicture}[baseline=(v.base)]
	\begin{feynman}[horizontal=p1 to v]
		\vertex (p1) [label=\({\scriptstyle f_{1i}}\)];
		\vertex [below right = of p1, blob, fill=gray] (v) {};
		\vertex [below left = of v,label=270:\({\scriptstyle \bar f_{2j}}\)] (p2);
		\vertex [right = of v,label=\({\scriptstyle V_3^c}\)] (p3);
		\diagram*{
			(p1) -- [fermion,momentum=\({\scriptstyle p_1}\)] (v)
			     -- [fermion,rmomentum'=\({\scriptstyle p_2}\)] (p2),
			(v) -- [photon,rmomentum'=\({\scriptstyle p_3}\)] (p3),
		};
	\end{feynman}
	\end{tikzpicture}
	}}}
	=
	g_{ f f V}^{ijc}\, (T^c)_{ji}\,
	\bar v_2^{(\rho_2)}(p_2)
	\slashed\!\epsilon_3^{\, \, (\rho_3)}
	u_1^{(\rho_1)}(p_1) ,
	\label{eq:Vff}
\end{align}
where all external states are on-shell $p_i^2 =m_i^2$, and $\rho_i$ label the polarization of the external fermion/vector boson. Here, \(a,b,c\) are adjoint gauge-group indices labeling the vector bosons, and \(f^{abc}\) are the corresponding structure constants. \(i,j\) are indices in the gauge-group representation carried by the fermions, \((T^c)_{ji}\) are the corresponding representation matrices, and $g$, $g_{ f f V}^{ijc}$ denote the coupling constants. 

Importantly, these minimal amplitudes satisfy the following off-shell \emph{current constraints} at the three-point level:
\begin{align}
 p_3^{\mu} A_{(VVV)\mu}^{(\rho_1\rho_2)}
 &=
 g f^{abc}\,
 (m_2^2-m_1^2)
 \bigl(\epsilon_1^{(\rho_1)}\cdot\epsilon_2^{(\rho_2)}\bigr), \label{cc:VVV}
 \\
 p_3^{\mu} A_{(f_i f_j V^c)\mu}^{(\rho_1\rho_2)}
 &=
 g_{ffV}^{ijc}(T^c)_{ji}\,
 \bar v_2\,(m_2-m_1)\,u_1 .
 \label{eq:ferm_dirac}
\end{align}
Contracting the polarization-stripped amplitude with $p_3^\mu$ reduces it to mass-dependent coefficients multiplying the external-state contractions shown above. These relations hold whether $p_3^\mu$ is on shell or off shell, provided that the other two legs are kept on shell.\footnote{For further discussion of off-shell current constraints for arbitrary massive spin-$s$ particles, see Ref.~\cite{Ema:2025qgd}.} In the special case in which the spin-1 leg is massless, $m_3=0$, gauge invariance requires the corresponding Ward identities,
\begin{align}
    \left.
    A_{V^aV^bV^c}^{(\rho_1\rho_2\rho_3)}
    \right|_{\epsilon_3^{(\rho_3)}\rightarrow p_3}
    =
    \left.
    A_{f_i f_j V^c}^{(\rho_1\rho_2\rho_3)}
    \right|_{\epsilon_3^{(\rho_3)}\rightarrow p_3}
    =0 ,
    \label{eq:three-point-Ward}
\end{align}
For nonzero couplings and generic external states, these identities imply $m_1=m_2$, as expected for two states coupled to the same massless gauge boson.

We now discuss three-point amplitudes involving longitudinal gauge bosons and their massless limits. It is convenient to decompose the longitudinal polarization vector as:
\begin{align}
    \epsilon_{L}^{\mu}(p)= \frac{p^{\mu}-2\eta^{\mu}}{m}, \qquad p^2 =m^2.
    \label{eq:long}
\end{align}
Because this expression contains an explicit factor of $1/m$, a generic three-point amplitude with longitudinal polarization can diverge in the $m \rightarrow 0$ limit. Given that $\eta^{\mu}= \frac{m^2}{2p \cdot n} n^{\mu}\sim \mathcal{O}(  m^2)$, the second term above has a smooth massless limit. Using Eq.~\eqref{eq:long}, we split the three-point amplitudes with the longitudinal external legs into scalar-like amplitudes coming from the $p^{\mu}/m$ terms, and separately $\eta$-dependent terms. The three-point $VVV$ amplitude with at least one longitudinal polarization is rewritten in the following form:

\begin{align}
	&A_{V^aV^bV^c}^{(L\rho_2 \rho_3)}
	=A_{\phi^a V^bV^c}^{(\rho_2 \rho_3)}+A_{V^aV^bV^c}^{(\eta_1 \rho_2 \rho_3)} \nonumber \\
    &\hspace{3.7em}=
	\frac{gf^{abc}}{m_1 } \Bigg\{\left[(m_3^2-m_2^2)(\epsilon_2^{(\rho_2)} \cdot \epsilon_3^{(\rho_3)})\right]+  \Bigl[- 2( \eta_1\cdot\epsilon_2^{(\rho_2)}) \,(p_1-p_2)\cdot\epsilon_3^{(\rho_3)} \nonumber \\
    &\hspace{6em}
	-2 (\epsilon_2^{(\rho_2)}\cdot\epsilon_3^{(\rho_3)}) \,\eta_1\cdot (p_2-p_3)-2(\epsilon_3^{(\rho_3)} \cdot \eta_1) \, (p_3-p_1)\cdot\epsilon_2^{(\rho_2)}\Bigr] \Bigg\},
	\label{eq:LVV}
	\\[1em]
    &A_{V^aV^bV^c}^{(LL \rho_3)}
	= A_{\phi^a \phi^b V^c}^{(\rho_3)}+ A_{V^a\phi^b V^c}^{( \eta_1 \rho_3)}+A_{\phi^a V^bV^c}^{( \eta_2 \rho_3)}\nonumber\\&\quad+ A_{V^aV^bV^c}^{( \eta_1 \eta_2 \rho_3)}\nonumber \\
	    &= \frac{gf^{abc}}{m_1 m_2} \Bigg\{\Bigl[{(m_3^2 -m_2^2-m_1^2)}\frac{p_2 -p_1}{2} \cdot \epsilon_3^{(\rho_3)}\Bigr] +\Bigl[ 2 (m_3^2 -m_1^2) (\eta_1 \cdot \epsilon_3^{(\rho_3)} )\Bigr] \nonumber\\&\quad+ \Bigl[- 2  (m_3^2 -m_2^2)(\eta_2 \cdot \epsilon_3^{(\rho_3)})\Bigr] \nonumber \\
    &+\Bigl[4 ( \eta_1\cdot\eta_2) (p_1-p_2)\cdot\epsilon_3^{(\rho_3)}
	+4 (\eta_2 \cdot\epsilon_3^{(\rho_3)}) \eta_1\cdot (p_2-p_3) + 4(\epsilon_3^{(\rho_3)} \cdot \eta_1)  (p_3-p_1)\cdot \eta_2 \Bigr] \Bigg \},
	\label{eq:LLV} \\
&A_{V^aV^bV^c}^{(LLL)}
=A_{V^a\phi^b\phi^c}^{(\eta_1)}
+
A_{\phi^aV^b\phi^c}^{(\eta_2)}
+
A_{\phi^a\phi^bV^c}^{(\eta_3)}
+
A_{V^aV^b\phi^c}^{(\eta_1\eta_2)}
+
A_{\phi^aV^bV^c}^{(\eta_2\eta_3)}
+
A_{V^a\phi^bV^c}^{(\eta_1\eta_3)}
+
A_{V^aV^bV^c}^{(\eta_1\eta_2\eta_3)}
\nonumber\\
&=
\frac{gf^{abc}}{m_1m_2m_3}
\Biggl\{
\Bigl[
m_1^2\,\eta_1\cdot(p_2-p_3)
\Bigr]
+
\Bigl[
m_2^2\,\eta_2\cdot(p_3-p_1)
\Bigr]
+
\Bigl[
m_3^2\,\eta_3\cdot(p_1-p_2)
\Bigr]
\nonumber\\
&+
4\Bigl[
(m_2^2-m_1^2)(\eta_1\cdot\eta_2)
+
(\eta_2\cdot p_3)\,
\eta_1\cdot(p_2-p_3)+
(\eta_1\cdot p_3)\,
\eta_2\cdot(p_3-p_1)
\Bigr]
\nonumber\\
&+
4\Bigl[
(m_3^2-m_2^2)(\eta_2\cdot\eta_3)
+
(\eta_3\cdot p_1)\,
\eta_2\cdot(p_3-p_1)
+
(\eta_2\cdot p_1)\,
\eta_3\cdot(p_1-p_2)
\Bigr]
\nonumber\\
&+
4\Bigl[
(m_1^2-m_3^2)(\eta_1\cdot\eta_3)
+
(\eta_3\cdot p_2)\,
\eta_1\cdot(p_2-p_3)
+
(\eta_1\cdot p_2)\,
\eta_3\cdot(p_1-p_2)
\Bigr]
\nonumber\\
&
-
8\Bigl[
(\eta_2\cdot\eta_3)\,
\eta_1\cdot(p_2-p_3)
+
(\eta_3\cdot\eta_1)\,
\eta_2\cdot(p_3-p_1)
+
(\eta_1\cdot\eta_2)\,
\eta_3\cdot(p_1-p_2)
\Bigr]
\Biggr\}.
\label{eq:LLL}
\end{align}
Eqs.~\eqref{eq:LVV}–\eqref{eq:LLL} represent the amplitudes with one, two, and three longitudinal external legs, respectively, with all remaining legs transverse.  On the left-hand side, each amplitude is labeled by the positions of its longitudinal legs, denoted by \(L\).  On the right side, the contribution obtained by replacing the longitudinal polarization on leg \(i\) with its \(p_i^\mu/m_i\) component is denoted by replacing \(V_i^a\) with \(\phi_i^a\). Contributions arising from the \(\eta_i^\mu\) component are explicitly indicated in the superscript. For example, in Eq.~\eqref{eq:LVV}, \(A_{\phi^a V^bV^c}^{(\rho_2\rho_3)}\) denotes the scalar-like contribution obtained from the \(p_1^\mu/m_1\) component, and \(A_{V^aV^bV^c}^{(\eta_1\rho_2\rho_3)}\) collects all terms containing  \(-2\eta_1^\mu/m_1\). Finally, we comment that the split into $\phi_i$ and $\eta_i$ pieces is not unique, as
$\eta_i\cdot p_i = m_i^2/2$ allows $\mathcal{O}(m_i^2)$ terms to be traded
between them.  

Similarly, 
for a longitudinally polarized vector, the gauge boson fermion amplitude becomes 
\begin{align} 
A_{f_i f_jV^c}^{(\rho_1\rho_2 L)}  &= A_{f_if_j\phi^c}^{(\rho_1\rho_2)} + A_{f_i f_j V^c}^{(\rho_1\rho_2\eta_3)}  \nonumber \\
&= \frac{g_{f fV}^{ijc}}{m_3}\,(T^c)_{ji} \bar v_2 \bigg[  \left(m_2-m_1\right)- 2\slashed\eta_3  \bigg] u_1 . 
\label{eq:LffV-expanded}
\end{align} 
With $\eta_i\sim \mathcal{O}(m_i^2)$ for small mass, we  see that  the three-point amplitudes in Eqs.~\eqref{eq:LVV}-\eqref{eq:LffV-expanded} are well defined in the simultaneous massless limit
\begin{align}
m_i=\tau \, \bar m_i,
\qquad
\tau\rightarrow 0,
\end{align}
with the mass ratios $\bar m_i/\bar m_j $ held fixed and the complex on-shell momenta continuously deformed so that momentum conservation and the on-shell conditions remain satisfied. This limit can be realized, e.g., if the three masses  are related to the same vev. By contrast, taking only one mass to zero, for example $m_3\rightarrow0$, while keeping $m_1$ and $m_2$ finite does not define a smooth limit unless $m_1=m_2$.

From the perspective of on-shell calculations, one may naively expect that constructing higher-point amplitudes using three-point amplitudes with a smooth massless limit will not be divergent. However, the four- and higher-point amplitudes are constructed from the \textit{shifted} three-point amplitudes and the latter may contain mass-singular terms.  For instance, the ALT shift deforms $\hat{\eta}^{\mu} = \eta^{\mu} + z r^{\mu}/2$ for a longitudinal external state, where the second term does not scale as $\mathcal{O}(m_i^2)$.
In the presence of longitudinal polarizations, we see from Eqs.~\eqref{eq:LVV}-\eqref{eq:LffV-expanded} that the $\eta$-dependent pieces generate mass-singular contributions after the shift. In Sec.~\ref{sec:ext_long}, we will use this observation to isolate mass-singular terms in higher-point amplitudes in different factorization channels.

\section{Off-shell continuation of internal states}
 \label{sec:off-shell}
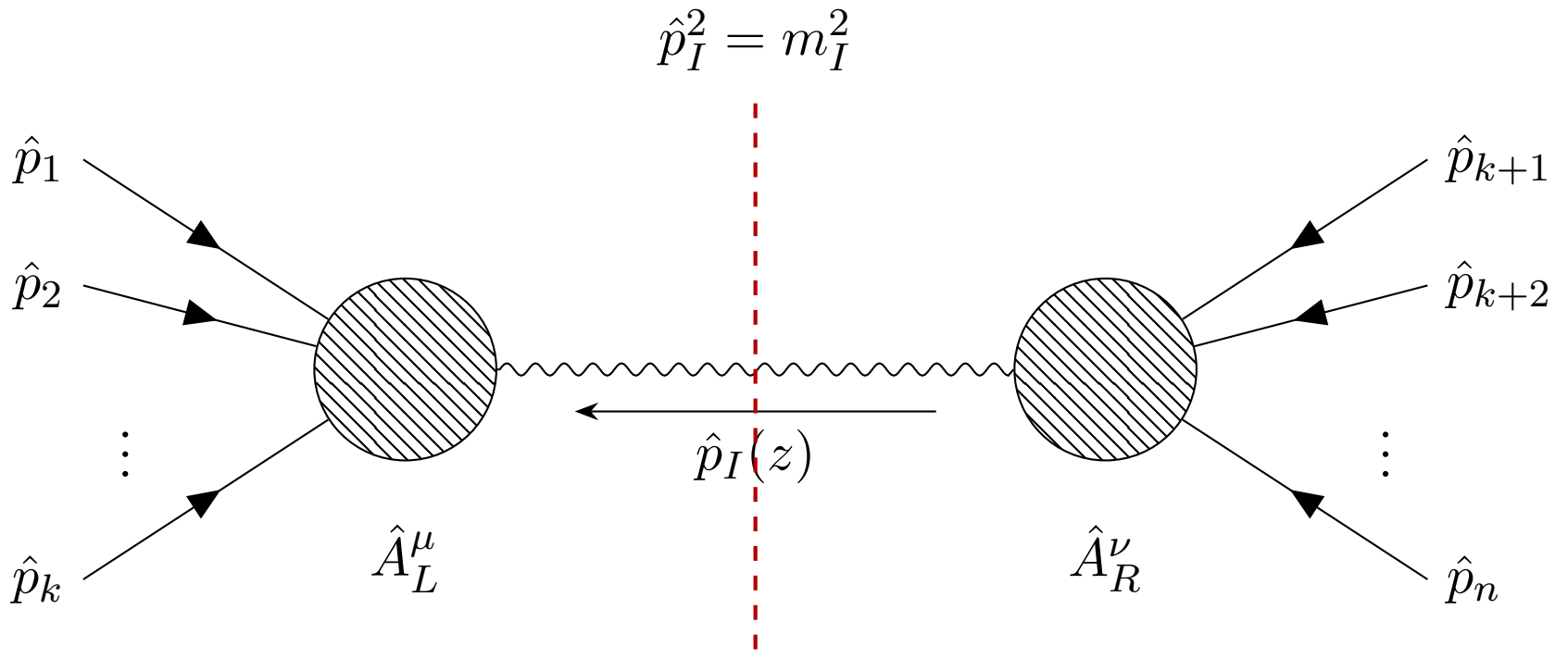
\begin{figure}[t]
\centering
\begin{tikzpicture}
  \begin{feynman}
    \vertex[blob, minimum size=1.3cm] (L) at (0,0) {};
    \vertex[blob, minimum size=1.3cm] (R) at (5,0) {};
    \vertex[label=180:\(\hat{p}_{1}\)] (i1) at (-2.3, 1.5);
    \vertex[label=180:\(\hat{p}_{2}\)] (i2) at (-2.3, 0.6);
    \vertex[label=180:\(\hat{p}_{k}\)] (i3) at (-2.3,-1.5);
    \vertex[label=0:\(\hat{p}_{k+1}\)] (f1) at (7.3, 1.5);
    \vertex[label=0:\(\hat{p}_{k+2}\)] (f2) at (7.3, 0.6);
    \vertex[label=0:\(\hat{p}_{n}\)]   (f3) at (7.3,-1.5);
    \diagram*{
      (i1) -- [fermion] (L),
      (i2) -- [fermion] (L),
      (i3) -- [fermion] (L),
      (L)  -- [boson, rmomentum'=\(\hat{p}_{I}(z)\)] (R),
      (f1) -- [fermion] (R),
      (f2) -- [fermion] (R),
      (f3) -- [fermion] (R),
    };
  \end{feynman}
  \node at (-2.0,-0.5) {\(\vdots\)};
  \node at ( 7.0,-0.5) {\(\vdots\)};
  \node at (0,-1.35) {\(\hat{A}_{L}^{\mu}\)};
  \node at (5,-1.35) {\(\hat{A}_{R}^{\nu}\)};
  \draw[dashed, thick, red!70!black] (2.5,-2.0) -- (2.5,2.0);
  \node[above] at (2.5,2.0) {\(\hat{p}_{I}^{2}=m_{I}^{2}\)};
\end{tikzpicture}
\caption{Factorization of the $n$-point amplitude $\hat A_{n}$ on the massive
vector channel $I$.}
\label{fig:factorization}
\end{figure}
In this section, we consider an $n$-point amplitude $A_n$ containing an internal massive spin-one exchange shown schematically in Fig. \ref{fig:factorization}. When the internal line momentum becomes on-shell, the amplitude factorizes into two lower-point on-shell amplitudes connected by a sum over the physical polarization states of the exchanged vector boson. We decompose the polarization sum into transverse and longitudinal contributions and study the two polarization contributions independently, and further determine how the projected amplitudes behave in the massless limits.

For a given factorization channel $I$, let
\begin{equation}
\hat p_I^{\mu}(z)=-\sum_{i\in L}\hat p_i^{\mu}(z) = -\sum_{i\in L}(p_i^{\mu}+ z r_i^{\mu})\equiv  p_I^{\mu} + z r_I^{\mu},
\end{equation}
where $\hat p_I^{\mu}$ denotes the shifted momentum carried by the exchanged vector boson, and $L$ denotes the set of legs on the left of the cut. 
The amplitude factorizes when the intermediate particle is on-shell: 
\begin{equation}
\hat p_I^2(z)-m_I^2
=
p_I^2-m_I^2
+
2z\,p_I\cdot r_I
+
z^2 r_I^2
=0 .
\label{eq:Delta-I}
\end{equation}
As is explained in Appendix~\ref{app:alt}, the contribution of the massive vector pole to the total amplitude is given by summing the residues at the poles. There are two situations depending on $r_I^2\neq0$ or $r_I^2=0$, and subsequently there are either two roots ($z_I^+$ and $z_I^-$), counted with multiplicity, or one root of Eq.~\eqref{eq:Delta-I}, provided $p_I\cdot r_I\ne0$ in the latter case.   In the absence of the boundary term, Eq.~\eqref{eq:alt-two-root} reduces to: 
\begin{align}
 \left. A_n\right|_{\rm vec\text{-}pole}^{\rm (two\text{-}root)}
 &=\frac{1}{(p_I^2-m_I^2)(z_I^+-z_I^-)}\Bigl[
 z_I^+\hat A_{L\mu}(z_I^-)\hat\Pi^{\mu\nu}(\hat p_I(z_I^-))\hat A_{R\nu}(z_I^-)
 \nonumber\\&\qquad
 -z_I^-\hat A_{L\mu}(z_I^+)\hat\Pi^{\mu\nu}(\hat p_I(z_I^+))\hat A_{R\nu}(z_I^+)\Bigr]
 \nonumber\\
 &=\frac{1}{p_I^2-m_I^2}\frac{z_I^+\mathcal N_I(z_I^-)-z_I^-\mathcal N_I(z_I^+)}{z_I^+-z_I^-},\qquad r_I^2\ne0.
 \label{eq:ALT-pole-two-root}
\end{align}
or
\begin{align}
\left. A_n\right|_{\rm vec\text{-}pole}^{\rm (single\text{-}root)} &= \frac{\mathcal{N}_I(z_I)}{p_I^2-m_I^2}, \qquad \text{when}\qquad r_I^2=0, \qquad p_I\cdot r_I\neq0 . 
\label{eq:ALT-pole-single-root} \end{align}
Here we have introduced a shorthand notation \footnote{For $\Pi_\rho^{\mu\nu}
=\epsilon_\rho^\mu\epsilon_{\bar\rho}^\nu$ in complex momenta, we use the analytic dual
$\epsilon_{\bar\rho}$, where $\bar+=-$, $\bar-=+$, and $\bar L=L$.
Thus the star in a real-momentum polarization sum is replaced by this dual
before applying the complex shift. No complex conjugation of $z$ is taken.}
\begin{equation}
\mathcal{N}_I(z)
\equiv
\hat A_{L\mu}(z)\,
\hat\Pi^{\mu\nu}\bigl(\hat p_I(z)\bigr)\,
\hat A_{R\nu}(z) .
\end{equation}
 for the numerator in channel \(I\). $\hat A_L^{\mu}(z)$ and $\hat A_R^{\nu}(z)$ denote the left and right factorized subamplitudes. The projectors 
\begin{equation}
\hat\Pi^{\mu\nu}(\hat p_I (z_I)) = \hat\Pi_+^{\mu\nu}+\hat\Pi_-^{\mu\nu}+\hat\Pi_L^{\mu\nu}
\end{equation}
are constructed from the transverse and longitudinal polarization vectors of the massive vector boson evaluated at the poles. This allows us to separate longitudinal and transverse contributions from the internal boson to the amplitude. Factorization occurs on the physical pole, and gauge-dependent contributions associated with off-shell propagators do not appear in the  residue. There is still a freedom in the choice of the  spin
quantization axis for the exchanged boson,
 which is specified by the null reference vector $n^\mu_I$ in Eq.~\eqref{eq:q_lambda_n}. 
 Different choices of $n^\mu_I$
mix the individual projectors $\hat{\Pi}_{\pm}^{\mu\nu}$ and
$\hat{\Pi}_{L}^{\mu\nu}$, and therefore would modify the decomposition into transverse and longitudinal contributions. 
Now, writing the shifted momentum as:
\begin{align}
\hat p_I^{\mu}   \equiv  \hat \lambda_I^{\mu}+ \frac{m_I^2}{2 \hat \lambda_I \cdot \hat n_I} \hat n_I^{\mu}= p_I^{\mu} + z r_I^{\mu},
\label{eq:hat p_I}
\end{align}
we can fix a reference vector \(\hat n^\mu_I\) such that all the \(z\)-dependence in \(\hat p_I^\mu\) is carried by \(\hat\lambda^\mu_I\). To achieve this, we require  that \(\hat n^\mu_I\) is independent of $z$ and satisfies \footnote{For shifts with $r_I^2 =0$, one choice for $r_I \cdot n_I =0$ is to set $n^{\mu}_I = r^{\mu}_I$, provided $p_I\cdot r_I\ne0$.}
\begin{align}
    r_I \cdot  n_I=0 .
    \label{eq:rn1}
\end{align} 
Namely, $n_I^{\mu}$ is orthogonal to the shift vector $r_I^{\mu}$. In this case, 
\begin{align}
\hat p_I\cdot n_I= \hat \lambda_I \cdot n_I =p_I\cdot n_I = \lambda_I \cdot n_I. 
\label{eq:n_I cond}
\end{align}
 
Thus $\hat\lambda_I^\mu(z)=p_I^\mu-\eta_I^\mu+z r_I^\mu$, where $\eta_I^\mu=m_I^2 n_I^\mu/(2p_I\cdot n_I)$ is independent of $z$ and $r_I^\mu=-\sum_{i\in L}r_i^\mu$ is the induced channel shift.

 The above relation   removes the $z$ dependence of the denominator of Eq.~\eqref{eq:hat p_I}. The existence of a null reference vector that satisfies Eq.~\eqref{eq:n_I cond}, together with the details of the null decomposition, is discussed in App.~\ref{app:null-decomposition}. The construction applies to shifts admitting a fixed null vector with $r_I\cdot n_I=0$ and $p_I\cdot n_I\ne0$; Appendix~\ref{app:null-existence} gives the existence condition.

 Due to  \eqref{eq:n_I cond}, the shifted transverse projector  
 is linear in $z$:
\begin{align} 
\hat \Pi_T^{\mu\nu}(z) = \hat \Pi_+^{\mu\nu}(z) +\hat \Pi_-^{\mu\nu}(z)  =\Pi_{T,0}^{\mu\nu} +z\,\Pi_{T,1}^{\mu\nu}, \label{eq:Tras_z} \end{align}
where
\begin{align}
 \Pi_{T,0}^{\mu\nu}
 &=-g^{\mu\nu}
   +\frac{2p_I^{(\mu}n_I^{\nu)}}{p_I\cdot n_I}
   -\frac{m_I^2 n_I^\mu n_I^\nu}{(p_I\cdot n_I)^2},
 \label{eq:tt}\\
 \Pi_{T,1}^{\mu\nu}
 &=\frac{2r_I^{(\mu}n_I^{\nu)}}{p_I\cdot n_I}.
\end{align}
writing $2a^{(\mu}b^{\nu)} = a^\mu b^\nu + a^\nu b^\mu$. Similarly, the longitudinal projector can be expanded up to quadratic order in $z$:
\begin{align}
\hat \Pi_L^{\mu\nu}(z) &= \Pi_{L,0}^{\mu\nu} +z\,\Pi_{L,1}^{\mu\nu} +z^2\,\Pi_{L,2}^{\mu\nu}, \label{eq:long_z}
\end{align} 
where 
\begin{align} \Pi_{L,0}^{\mu\nu} &= \frac{p_I^\mu p_I^\nu}{m_I^2} -\frac{2p_I^{(\mu}n_I^{\nu)}}{p_I\cdot n_I} +\frac{m_I^2 n_I^\mu n_I^\nu} {(p_I\cdot n_I)^2}, \\ \Pi_{L,1}^{\mu\nu} &= \frac{2p_I^{(\mu}r_I^{\nu)}}{m_I^2} -\frac{2r_I^{(\mu}n_I^{\nu)}}{p_I\cdot n_I}, \\ \Pi_{L,2}^{\mu\nu} &= \frac{r_I^\mu r_I^\nu}{m_I^2}.
\end{align}
Another property of  Eq. \eqref{eq:n_I cond} is that the denominator $p_I\cdot n_I$  in the projectors  is independent of $z$. As a consequence, the spurious $z$-poles are avoided. In the following, we will study the projected amplitudes within this construction, through several examples.


\subsection{\texorpdfstring{$f  f f f$}{ffff} case}
We begin with the four-fermion amplitude \[
A_4\bigl(f_i(p_1),f_j(p_2),f_k(p_3), f_l(p_4)\bigr),
\]
with massive vector-boson exchange. All external momenta are taken to be incoming,
$\sum_{a=1}^{4}p_a=0$, and we label the fermions $f_i,\,f_j,\,f_k,\, f_l$ by
$1,2,3,4$, with helicities $\rho_1,\dots,\rho_4$ and masses $m_1,\dots,m_4$,
which we take to be nonzero and unequal in general. Here $i,j,k,l$ are indices in
the gauge-group representation carried by the fermions, in the same convention as
Eq.~\eqref{eq:Vff}. To illustrate the
decomposition of the internal vector boson we study only the $s$-channel
factorization, in which the exchanged momentum is
\begin{align}
\hat p_I=-(\hat p_1+\hat p_2)=\hat p_3+\hat p_4 ,
\label{eq:s-mom}
\end{align}
Focusing on the numerator of the residue in  \eqref{eq:ALT-pole-two-root} and  \eqref{eq:ALT-pole-single-root} after the shift, we find the projection onto transverse and longitudinal components of the internal polarization to be given by
\begin{align}
   & \mathcal{N}_{I,T} (z_I)= \hat A_{( f_if_j V^c ) \mu }^{(\rho_1 \rho_2)}\hat \Pi_T^{\mu\nu} \hat  A_{( f_k  f_l V^c) \nu}^{(\rho_3 \rho_4)} 
  = A_{( f_i  f_j V^c ) \mu }^{(\rho_1 \rho_2)} \Pi_{T,0}^{\mu\nu}   A_{( f_k  f_l V^c) \nu}^{(\rho_3 \rho_4)} \label{eq:Tffff}\\
  &\mathcal{N}_{I,L} (z_I)= \hat A_{(f_if_j V^c) \mu }^{(\rho_1 \rho_2)} \hat\Pi_L^{\mu\nu}
  {\hat A_{(f_k f_l V^c) \nu}^{(\rho_3 \rho_4)}}=  A_{(f_i f_j V^c) \mu }^{(\rho_1 \rho_2)} \Pi_{L,0}^{\mu\nu}
  {A_{(f_k f_l V^c) \nu}^{(\rho_3 \rho_4)}}
     \label{eq:Lffff}
\end{align}
where $A_{(f_i f_j V^c) \mu }^{(\rho_1 \rho_2)}, \, A_{(f_k  f_l V^c) \nu }^{(\rho_3 \rho_4)}$ are the three-point amplitudes defined in Eq. \eqref{eq:Vff} with the vector polarization stripped off; $z_I$ denotes either of the pole solutions $\hat p_I^2(z_I)=m_I^2$. The adjoint index $c$ labels the species of the exchanged
vector, giving the factor
$(T^c)_{ji}(T^c)_{lk}$, which we leave implicit.

The last equalities above show that all the $z_I$-dependent terms that make the amplitude on-shell simply vanish in the contraction, and the amplitude is expressed in terms of the unshifted off-shell variables in $\Pi_{T,0}^{\mu\nu}$ and $\Pi_{L,0}^{\mu\nu}$. To derive this, we used the fact that these particular three-point amplitudes are independent of $z_I$, i.e.~$ \hat A_{( f_if_j V^c ) \mu }^{(\rho_1 \rho_2)}= A_{( f_if_j V^c ) \mu }^{(\rho_1 \rho_2)}$, since the external spinors are not deformed by ALT shift. The $z_I$-dependent terms in $\hat \Pi^{\mu \nu}$ vanish after contracting with the subamplitudes, due to the relation 
\begin{align}
    r_I^{\mu} \, A_{(f_if_j V^c) \mu  }^{(\rho_1 \rho_2)}=r_I^{\mu} A_{(f_k f_l V^c)  \mu}^{(\rho_3 \rho_4)}=0 \,, \qquad
r_I^{\mu}=-r_1^{\mu}-r_2^{\mu}.
    \label{eq:rIcdotffv}
\end{align}
This is a consequence of the minimal structure of the three-point amplitude satisfying the off-shell current constraint
\begin{align}
    \hat p_I^{\mu} \, \hat A_{(f_i f_j V^c) \mu  }^{(\rho_1 \rho_2)} \, = p_I^{\mu} \, A_{(f_i f_j V^c) \mu  }^{(\rho_1 \rho_2)} \, =m_I A_{f_if_j\phi^c}^{(\rho_1\rho_2)} 
    .\label{eq:ferm_dirac1}
\end{align} We then recover \eqref{eq:rIcdotffv} after using  $\hat p_I^{\mu} = p_I^{\mu}+ z_I r^\mu_I$.
The same argument, with $(1,2) \to (3,4)$, gives $r_I \cdot A_{(f_k f_l V^c)}^{(\rho_3 \rho_4)} = 0$. 

Using Eq.~\eqref{eq:ferm_dirac} which holds for off-shell $p_I^{\mu}$, we can also rewrite \eqref{eq:Lffff} as: 
\begin{align}
 \mathcal{N}_{I,L} (z_I) &= - A_{f_if_j\phi^c}^{(\rho_1\rho_2)}  A_{f_kf_l\phi^c}^{(\rho_3\rho_4)}  +A_{f_if_j\phi^c}^{(\rho_1\rho_2)}A_{f_kf_l V^c}^{(\rho_3\rho_4\eta_I)} - A_{f_if_jV^c}^{(\rho_1\rho_2 \eta_I)}A_{f_kf_l\phi^c}^{(\rho_3\rho_4)} +A_{f_if_jV^c}^{(\rho_1\rho_2 \eta_I)}  A_{f_kf_l V^c}^{(\rho_3\rho_4\eta_I)}.
  \label{eq:ffffL_explicit}
\end{align}
where $\eta_I^{\mu} = \frac{m_I^2}{2p_I\cdot n_I}n_I^{\mu}$. Both $\eta_I$ insertions use this same routed vector, including in the right subamplitude whose incoming vector momentum is $-p_I$. Thus the longitudinal contraction is $(A_{\phi,L}+A_{\eta,L})(-A_{\phi,R}+A_{\eta,R})$, where $A_{\eta,L/R}=-2\eta_I\cdot A_{L/R}/m_I$. The contribution of the $s$-channel pole to the amplitude is computed after plugging Eq.~\eqref{eq:Tffff}-\eqref{eq:Lffff} into Eq.~\eqref{eq:ALT-pole-two-root}:
\begin{align}
A_4^{(s)}=\frac{\mathcal{N}_{I,T}+\mathcal{N}_{I,L}}{p_I^2-m_I^2},
 \label{eq:num}
\end{align}
which is the contraction of the two fermionic currents with the numerator of the massive-vector propagator in unitary gauge. We conclude by noting that Eq. \eqref{eq:Tffff}  and Eq. \eqref{eq:ffffL_explicit} provide 
a decomposition of the internal boson propagator into transverse and longitudinal modes with off-shell momentum.
This decomposition is inspired by the ALT shift, and the current constraint played a central role at several points: it removed the
 $z_I$-dependent terms from both projections and traded the explicit
$1/m_I^2$ of $\Pi_{L,0}^{\mu\nu}$ for a product of scalar-like and
$\eta_I$-dependent amplitudes. Furthermore, by removing the $1/m_I^2$ from the propagator, it improves the large-$z$ scaling of the full amplitude.

\subsection{\texorpdfstring{$VVff$}{VVff} case}
Next we consider the vector-exchange contribution to the four-point amplitude
\[
A_4\bigl(V^a(p_1),\,V^b(p_2),\,f_i(p_3),\,f_j(p_4)\bigr),
\]
labeling the legs $1,2,3,4$ with helicities $\rho_1,\dots,\rho_4$ and masses
$m_1,\dots,m_4$. Here $a,b$ are adjoint gauge-group indices carried by the external vectors and $i,j$
are indices in the representation carried by the fermions. We only study the $s$-channel factorization with massive vector exchange. For $s$-channel, the left subamplitude is the amputated $VVV$ amplitude in
Eq.~\eqref{eq:VVV} and the right subamplitude the fermionic current in
Eq.~\eqref{eq:Vff}, so that the channel carries the common group-theory factor
$f^{abc}(T^c)_{ji}$, where $c$ is the adjoint index of the exchanged vector. 
For the transverse projection, we focus on the numerator of Eq.~\eqref{eq:ALT-pole-two-root}. Below, $\stackrel{\rm sum}{=}$ denotes equality after the weighted two-root sum in that equation, rather than equality at either individual pole.
\begin{align}
\mathcal{N}_{I,T}(z_I)
&= \hat A_{(V^aV^bV^c)\mu}^{(\rho_1\rho_2)}\,
   \hat\Pi_T^{\mu\nu}\,
   \hat A_{(f_if_jV^c)\nu}^{(\rho_3\rho_4)}
\nonumber\\
&= A_{(V^aV^bV^c)\mu}^{(\rho_1\rho_2)}\,\Pi_{T,0}^{\mu\nu}\,
   A_{(f_if_jV^c)\nu}^{(\rho_3\rho_4)}
 + z_I\,A_{(V^aV^bV^c)\mu}^{(\rho_1\rho_2)}\,\Pi_{T,1}^{\mu\nu}\,
   A_{(f_if_jV^c)\nu}^{(\rho_3\rho_4)}
\nonumber\\
&\quad
 + z_I\,A_{(V^aV^bV^c)\mu}^{r,(\rho_1\rho_2)}\,\Pi_{T,0}^{\mu\nu}\,
   A_{(f_if_jV^c)\nu}^{(\rho_3\rho_4)}
 + z_I^2\,A_{(V^aV^bV^c)\mu}^{r,(\rho_1\rho_2)}\,\Pi_{T,1}^{\mu\nu}\,
   A_{(f_if_jV^c)\nu}^{(\rho_3\rho_4)}
\nonumber\\
&\stackrel{\rm sum}{=} A_{(V^aV^bV^c)\mu}^{(\rho_1\rho_2)}\,\Pi_{T,0}^{\mu\nu}\,
   A_{(f_if_jV^c)\nu}^{(\rho_3\rho_4)} ,
\label{eq:TVVff}
\end{align}
where 
\begin{align}
A_{(V^aV^bV^c)\mu}^{r,(\rho_1\rho_2)}
&= gf^{abc}\Bigl[
 (\epsilon_1^{(\rho_1)}\!\cdot\!\epsilon_2^{(\rho_2)})(r_1-r_2)_\mu
 + 2(\epsilon_1^{(\rho_1)}\!\cdot\!r_2)\,\epsilon_{2\mu}^{(\rho_2)}
 - 2(\epsilon_2^{(\rho_2)}\!\cdot\!r_1)\,\epsilon_{1\mu}^{(\rho_1)}
\Bigr] .
\label{eq:VVff-VVV-r}
\end{align}
The $z_I^2$ term in the amplitude vanishes due to Eq. \eqref{eq:rIcdotffv} and Ward identity for the shifted $VVV$ amplitude
\begin{align}
r_I^{\mu}\,A_{(V^aV^bV^c)\mu}^{r,(\rho_1\rho_2)} = 0 .
\label{eq:shifted_ward}
\end{align}
We drop the linear order terms in $z_I$ from the expression as they cancel after summing over the two poles $z_I^{\pm}$ in Eq.~\eqref{eq:ALT-pole-two-root} (see Eq. \eqref{eq:sum_poles} for more details). Therefore, we can write the amplitude in terms of   $\Pi_{T,0}^{\mu\nu}$ only.

For longitudinal projection, after dropping terms linear in $z_I^{\pm}$, we get:
\begin{align}
\mathcal{N}_{I,L}(z_I)
&= \hat A_{(V^aV^bV^c)\mu}^{(\rho_1\rho_2)}\,\hat\Pi_L^{\mu\nu}\,
   \hat A_{(f_if_jV^c)\nu}^{(\rho_3\rho_4)}
 \stackrel{\rm sum}{=} A_{(V^aV^bV^c)\mu}^{(\rho_1\rho_2)}\,\Pi_{L,0}^{\mu\nu}\,
   A_{(f_if_jV^c)\nu}^{(\rho_3\rho_4)}
\nonumber\\[0.4em]
&= \Bigl(A_{\phi^cV^aV^b}^{(\rho_1\rho_2)}
        + A_{V^cV^aV^b}^{(\eta_I\rho_1\rho_2)}\Bigr)
   \Bigl(-A_{f_if_j\phi^c}^{(\rho_3\rho_4)}
        + A_{f_if_jV^c}^{(\rho_3\rho_4\eta_I)}\Bigr).
\label{eq:VVffL}
\end{align}
The last equality removes   $1/m_I^2$ from the internal longitudinal mode. Similarly to the previous example, this is achieved due to the $VVV$ interaction satisfying the current constraint:
\begin{align}
   \hat p_I^{\mu} \hat A_{(VVV)\mu }^{(\rho_1 \rho_2 )} = p_I^{\mu}  A_{(VVV)\mu }^{(\rho_1 \rho_2 )}= m_I A_{ \phi VV }^{(\rho_1 \rho_2)}.
   \label{eq:ward}
\end{align}
 where $A_{ \phi VV }^{(\rho_1 \rho_2)}$ is defined in Eq. \eqref{eq:LVV}. The relation holds for off-shell internal momentum $p_I^{\mu}$, and the split into transverse and longitudinal parts in Eqs.~\eqref{eq:TVVff} and \eqref{eq:VVffL} is expressed in terms of the
unshifted variables.

Let us now study the massless limit of the internal vector boson, assuming that the external
vector polarizations are transverse. A   smooth  massless limit requires the absence of $1/m_I^2$ terms. Furthermore, we require the decoupling of the longitudinal mode
of the internal vector, and so  the surviving transverse
contribution must be independent of the choice of spin quantization axis $n_I^\mu$. 

As $m_I\to0$ at fixed nonzero $p_I\cdot n_I$, the term proportional to
$m_I^2n_I^\mu n_I^\nu$ in $\Pi_{T,0}^{\mu\nu}$ vanishes. The remaining
$n_I$-dependent terms drop out when the two currents are conserved,
$p_I^\mu A_{(V^aV^bV^c)\mu}^{(\rho_1\rho_2)}
= p_I^\nu A_{(f_if_jV^c)\nu}^{(\rho_3\rho_4)} = 0$.
 In the limit $m_I\to0$ with the external masses held fixed, current conservation requires $m_1=m_2$ and
$m_3=m_4$. In that case,   the longitudinal contribution given in 
Eq.~\eqref{eq:VVffL} vanishes. Only the transverse contribution
of Eq.~\eqref{eq:TVVff} remains and is independent of $n_I^\mu$.

If instead $m_1\neq m_2$ or $m_3\neq m_4$, taking only $m_I\to0$ does not give a
  smooth massless gauge-boson limit. A smooth limit can be obtained by sending the external masses that
violate current conservation to zero along with $m_I$,
\begin{equation}
m_n(\tau)=\tau\,\overline m_n ,
\qquad
n\in\{1,2,3,4,I\},
\label{eq:massless_limit}
\end{equation}
at fixed reference masses $\overline m_n$ (and hence fixed mass ratios), followed by $\tau\to0$. In this limit
$A_{\phi^cV^aV^b}^{(\rho_1\rho_2)}=\mathcal{O}(\tau)$, while
$A_{f_if_j\phi^c}^{(\rho_3\rho_4)}=\mathcal{O}(1)$, since the ratio
$(m_3-m_4)/m_I$ is held fixed; their product, and hence the leading term of
$\mathcal{N}_{I,L}$, vanishes as $\mathcal{O}(\tau)$. The mixed and $\eta_I\eta_I$ terms
vanish as $\mathcal{O}(\tau)$ and $\mathcal{O}(\tau^2)$ respectively, provided the external
vectors are transversely polarized. 


\subsection{\texorpdfstring{$VVVV$}{VVVV} case}
We now consider the four-vector amplitude
\[
A_4\bigl(V^a(p_1),\,V^b(p_2),\,V^d(p_3),\,V^e(p_4)\bigr),
\]
labeling the legs $1,2,3,4$ with helicities $\rho_1,\dots,\rho_4$ and masses
$m_1,\dots,m_4$. We take the external vectors to be transversely polarized, so that $r_i^\mu\propto\epsilon_i^\mu$ in the contact-term reduction below. We study only the
$s$-channel exchange of a massive vector with adjoint index $c$, so that both
the left and right subamplitudes are amputated $VVV$ amplitudes of Eq.~\eqref{eq:VVV} and
the channel carries the group-theory factor $f^{abc}f^{cde}$ which is left implicit below. 

Unlike
the previous two cases, each subamplitude carries an $\mathcal{O}(z)$ piece as given  in  Eq.~\eqref{eq:VVff-VVV-r}, resulting in  four-point contact contributions \cite{Ema:2024rss}. The transverse
projection of the numerator in Eq.~\eqref{eq:ALT-pole-two-root} is
\begin{align}
\mathcal{N}_{I,T}(z_I)
&= \hat A_{(V^aV^bV^c)\mu}^{(\rho_1\rho_2)}\,\hat\Pi_T^{\mu\nu}\,
   \hat A_{(V^cV^dV^e)\nu}^{(\rho_3\rho_4)}
 \stackrel{\rm sum}{=} \mathcal{N}_{I,T}^{(0)} + z_I^2\,\mathcal{N}_{I,T}^{(2)} .
\end{align}
where
\begin{align}
\mathcal{N}_{I,T}^{(0)} &= A_{(V^aV^bV^c)\mu}^{(\rho_1\rho_2)}\,\Pi_{T,0}^{\mu\nu}\,
   A_{(V^cV^dV^e)\nu}^{(\rho_3\rho_4)} ,
\label{eq:VVVV-T0}\\
\mathcal{N}_{I,T}^{(2)} &= A_{(V^aV^bV^c)\mu}^{r,(\rho_1\rho_2)}\,\Pi_{T,0}^{\mu\nu}\,
   A_{(V^cV^dV^e)\nu}^{r,(\rho_3\rho_4)}
 + A_{(V^aV^bV^c)\mu}^{r,(\rho_1\rho_2)}\,\Pi_{T,1}^{\mu\nu}\,
   A_{(V^cV^dV^e)\nu}^{(\rho_3\rho_4)}
\nonumber\\
&\hspace{2em}
 + A_{(V^aV^bV^c)\mu}^{(\rho_1\rho_2)}\,\Pi_{T,1}^{\mu\nu}\,
   A_{(V^cV^dV^e)\nu}^{r,(\rho_3\rho_4)} ,
\label{eq:VVVV-T2-unsimplified}
\end{align}
The terms linear in $z_I$ cancel in the weighted sum over the two factorization roots, while the cubic term vanishes by Eq.~\eqref{eq:shifted_ward}.  Using Eqs.~\eqref{eq:rn1} and \eqref{eq:ward}, we can further simplify 
\begin{align}
\mathcal{N}_{I,T}^{(2)} = -A_{(V^aV^bV^c)}^{r,(\rho_1\rho_2)}\!\cdot\!A_{(V^cV^dV^e)}^{r,(\rho_3\rho_4)}
- \frac{m_I^2}{(p_I\!\cdot\!n_I)^2}
  \bigl(n_I\!\cdot\!A_{(V^aV^bV^c)}^{r,(\rho_1\rho_2)}\bigr)
  \bigl(n_I\!\cdot\!A_{(V^cV^dV^e)}^{r,(\rho_3\rho_4)}\bigr) .
\label{eq:VVVV-T2}
\end{align}
For the \(VVVV\) amplitude, after summing over the two poles, the transverse part of the \(s\)-channel contribution becomes 
\begin{align}
A_{4,T}^{(s)}
&= \frac{\mathcal{N}_{I,T}^{(0)} - z_I^+z_I^-\,\mathcal{N}_{I,T}^{(2)}}{p_I^2-m_I^2}
 = \frac{A_{(V^aV^bV^c)\mu}^{(\rho_1\rho_2)}\,\Pi_{T,0}^{\mu\nu}\,
         A_{(V^cV^dV^e)\nu}^{(\rho_3\rho_4)}}{p_I^2-m_I^2}
 - \frac{\mathcal{N}_{I,T}^{(2)}}{r_I^2} .
\label{eq:VVVV-T-final}
\end{align}
The second term above includes the  contact term of the four-point amplitude.  More precisely, it is shown \cite{Ema:2024rss} that the contact term arises from the   first term in  $\mathcal{N}_{I,T}^{(2)}$  after using $r_i^{\mu} \propto \epsilon_i^{\mu}$. The reference-vector-dependent second term in $\mathcal{N}_{I,T}^{(2)}$ cancels against $\mathcal{N}_{I,L}^{(2)}$ when the transverse and longitudinal contributions are combined, as shown below. Note that this term is proportional to
$m_I^2$ at fixed $p_I\cdot n_I$ and vanishes as $m_I\to0$. Together with
current conservation of the two subamplitudes (i.e.\ $m_1=m_2$, $m_3=m_4$, or the
simultaneous limit of Eq.~\eqref{eq:massless_limit}), the transverse projection
then reduces to that of a massless vector exchange.   

 For longitudinal projection
 \begin{align}
\mathcal{N}_{I,L}(z_I)
&= \hat A_{(V^aV^bV^c)\mu}^{(\rho_1\rho_2)}\,\hat\Pi_L^{\mu\nu}\,
   \hat A_{(V^cV^dV^e)\nu}^{(\rho_3\rho_4)}
 \stackrel{\rm sum}{=} \mathcal{N}_{I,L}^{(0)} + z_I^2\,\mathcal{N}_{I,L}^{(2)} ,
\end{align}
 where 
\begin{align}
\mathcal{N}_{I,L}^{(0)}
&= A_{(V^aV^bV^c)\mu}^{(\rho_1\rho_2)}\,\Pi_{L,0}^{\mu\nu}\,
   A_{(V^cV^dV^e)\nu}^{(\rho_3\rho_4)}
 = \Bigl(A_{\phi^cV^aV^b}^{(\rho_1\rho_2)}
        + A_{V^cV^aV^b}^{(\eta_I\rho_1\rho_2)}\Bigr)
   \Bigl(-A_{\phi^cV^dV^e}^{(\rho_3\rho_4)}
        + A_{V^cV^dV^e}^{(\eta_I\rho_3\rho_4)}\Bigr),
\label{eq:VVVV_L}\\[0.4em]
\mathcal{N}_{I,L}^{(2)}
&= \frac{m_I^2}{(p_I\!\cdot\!n_I)^2}
   \bigl(n_I\!\cdot\!A_{(V^aV^bV^c)}^{r,(\rho_1\rho_2)}\bigr)
   \bigl(n_I\!\cdot\!A_{(V^cV^dV^e)}^{r,(\rho_3\rho_4)}\bigr) .
\label{eq:VVVV-L2}
\end{align} 
All terms of cubic and quartic order in \(z_I\) vanish as a consequence of Eq.~\eqref{eq:ward}. Therefore, 
\begin{align}
A_{4,L}^{(s)}
&= \frac{\mathcal{N}_{I,L}^{(0)} - z_I^+z_I^-\,\mathcal{N}_{I,L}^{(2)}}{p_I^2-m_I^2}
 = \frac{A_{(V^aV^bV^c)\mu}^{(\rho_1\rho_2)}\,\Pi_{L,0}^{\mu\nu}\,
         A_{(V^cV^dV^e)\nu}^{(\rho_3\rho_4)}}{p_I^2-m_I^2}
 - \frac{\mathcal{N}_{I,L}^{(2)}}{r_I^2} .
\label{eq:VVVV-L-final}
\end{align}
 Adding Eqs.~\eqref{eq:VVVV-T-final} and \eqref{eq:VVVV-L-final}, the reference-vector-dependent quadratic terms cancel:
\begin{align}
\mathcal{N}_{I,T}^{(2)} + \mathcal{N}_{I,L}^{(2)}
= -A_{(V^aV^bV^c)}^{r,(\rho_1\rho_2)}\!\cdot\!A_{(V^cV^dV^e)}^{r,(\rho_3\rho_4)} .
\label{eq:VVVV-N2-sum}
\end{align}
 The complete \(s\)-channel vector-exchange contribution can therefore be written as 
 \begin{align}
A_{4,\mathrm{vec}}^{(s)}
&= \frac{A_{(V^aV^bV^c)\mu}^{(\rho_1\rho_2)}
         \bigl(\Pi_{T,0}^{\mu\nu}+\Pi_{L,0}^{\mu\nu}\bigr)
         A_{(V^cV^dV^e)\nu}^{(\rho_3\rho_4)}}{p_I^2-m_I^2}
 + A_{4}^{(s),\mathrm{contact}} ,
\label{eq:VVVV-total-s}\\
A_{4}^{(s),\mathrm{contact}}
&= \frac{A_{(V^aV^bV^c)}^{r,(\rho_1\rho_2)}\!\cdot\!
          A_{(V^cV^dV^e)}^{r,(\rho_3\rho_4)}}{r_I^2}
\nonumber\\
&= g^2f^{abc}f^{cde}
\Big[
(\epsilon_1\!\cdot\!\epsilon_4)
 (\epsilon_2\!\cdot\!\epsilon_3)
-
(\epsilon_1\!\cdot\!\epsilon_3)
 (\epsilon_2\!\cdot\!\epsilon_4)
\Big].
\label{eq:VVVV-contact}
\end{align}
The first term is the ordinary massive-vector pole contribution, where we project onto transverse and longitudinal components of the internal vector, with the longitudinal projection further given by  Eq.~\eqref{eq:VVVV_L}. 
The last equality in Eq.~\eqref{eq:VVVV-contact} uses $r_i^\mu=c_i\epsilon_i^\mu$ for all four external vectors and $\sum_i r_i=0$. It is the contact contribution assigned to this factorization channel; the full quartic interaction is obtained after summing the channel contributions with their group factors, as illustrated for $WWWW$ in Ref.~\cite{Ema:2024rss}.

The massless limit of $VVVV$ case follows similarly to the $VVff$ case. Taking only $m_I\to0$, we should require
$m_1=m_2$ and $m_3=m_4$ for current conservation, so that this massless limit is smooth. For unequal
external masses, if all the masses scale proportionally as in
Eq.~\eqref{eq:massless_limit}, there is also a smooth massless limit. For transversely polarized external vectors, $\mathcal{N}_{I,T}^{(0)}=\mathcal{O}(\tau^0)$ with reference-vector dependence vanishing in the limit, and the contact term is $\mathcal{O}(\tau^0)$, while $\mathcal{N}_{I,L}^{(0)}=\mathcal{O}(\tau^2)$.

\section{Longitudinal external states and mass singularities}
\label{sec:ext_long}
We now examine mass singularities associated with external longitudinal vectors. Using the $VVff$ amplitude as an example, we show how the shift-dependent terms in the polarization decomposition isolate potentially singular contact contributions and make their cancellation explicit.
\subsection{ \texorpdfstring{$VVf  f$}{VVff} case}
We consider the four-point amplitude
\begin{align}
A_4\bigl(V_1^a(p_1),\,V_2^b(p_2),\,f_i(p_3),\, f_j(p_4)\bigr),
\end{align}
 Here $a,b$ are adjoint labels for
the external vectors and $i,j$ are representation indices for the fermions,
with masses $m_1,\ldots,m_4$. Legs $1$ and $2$ are longitudinally polarized, while legs $3$ and $4$
carry helicities $\rho_3,\rho_4$. The amplitude factorizes on
three channels: $s$-channel exchange of a vector $V^c$ of mass $m_I$, and $t$- and
$u$-channel exchanges of fermions of mass $m_F^t$ and $m_F^u$.
For the $s$-channel factorization, the internal momentum is defined to be
\begin{align} 
\hat p_s^{\mu} &= -\left(\hat p_1^{\mu}+\hat p_2^{\mu}\right) = \hat p_3^{\mu}+\hat p_4^{\mu}, \label{eq:s-channel-momentum} \end{align}
 and the factorization poles are determined by 
$ \hat p_s^2(z_s^\pm)-m_I^2 = 0 $, with $m_I$ being the intermediate particle mass. Using the amputated three-point amplitude in Eq.~\eqref{eq:LLV}, and replacing $m_3$ by $m_I$, we rewrite the $VVV$ left subamplitude: \footnote{At the factorization pole, $\hat A_{(V^aV^bV^c)\mu}^{(LL)}$ is defined up to terms proportional to $\hat p_{s\mu}$, which vanish after contracting with $\hat \Pi_T^{\mu \nu}(\hat p_s(z_s))+\hat\Pi_L^{\mu \nu}(\hat p_s(z_s))$.}
\begin{align}
\hat A_{(V^aV^bV^c)\mu}^{(LL)}
&= \hat A_{(\phi^a\phi^bV^c)\mu}
+ \hat A_{(V^a\phi^bV^c)\mu}^{(\hat\eta_1)}
+ \hat A_{(\phi^aV^bV^c)\mu}^{(\hat\eta_2)}
+ \hat A_{(V^aV^bV^c)\mu}^{(\hat\eta_1\hat\eta_2)}.
\end{align}
The numerator of massive vector exchange contribution at the pole is given by
\begin{align}
\mathcal{N}_s(z_s)
&= \hat A_{(V^aV^bV^c)\mu}^{(LL)}
   \bigl(\hat\Pi_+^{\mu\nu}+\hat\Pi_-^{\mu\nu}+\hat\Pi_L^{\mu\nu}\bigr)
   \hat A_{(f_if_jV^c)\nu}^{(\rho_3\rho_4)}
\nonumber\\[0.3em]
&= -\,\hat A_{(V^aV^bV^c)}^{(LL)\,\mu}\,
      \hat A_{(f_if_jV^c)\mu}^{(\rho_3\rho_4)}
   +\frac{1}{m_I^2}
    \bigl(\hat p_s\!\cdot\!\hat A_{(V^aV^bV^c)}^{(LL)}\bigr)
    \bigl(\hat p_s\!\cdot\!\hat A_{(f_if_jV^c)}^{(\rho_3\rho_4)}\bigr) ,
\label{eq:VLVL-Ns-split}
\end{align}
where the two contractions are fixed by the current constraints
Eq. \eqref{eq:ward} and Eq. \eqref{eq:ferm_dirac},
\begin{align}
\hat p_s\!\cdot\!\hat A_{(f_if_jV^c)}^{(\rho_3\rho_4)}
&= g_{ffV}^{ijc}(T^c)_{ji}\,\left(m_3-m_4\right)\,
   \overline v_4^{(\rho_4)}u_3^{(\rho_3)} ,
\label{eq:VLVL-ps-right}
\\[0.4em]
\hat p_s\!\cdot\!\hat A_{(V^aV^bV^c)}^{(LL)}
&= \frac{gf^{abc}}{m_1m_2}
\bigg[
\frac{\left(m_2^2-m_1^2\right)\left(m_I^2-m_1^2-m_2^2\right)}{2}
+ 4\left(m_2^2-m_1^2\right)\bigl(\hat\eta_1\!\cdot\!\hat\eta_2\bigr)
\nonumber\\
&\hspace{5.2em}
+ 2\left(m_I^2+m_1^2-m_2^2\right)\bigl(\hat p_1\!\cdot\!\hat\eta_2\bigr)
- 2\left(m_I^2-m_1^2+m_2^2\right)\bigl(\hat\eta_1\!\cdot\!\hat p_2\bigr)
\bigg] .
\label{eq:VLVL-ps-left}
\end{align}
Expanding in the shift parameter,
\begin{align}
\mathcal{N}_s(z_s)
= \mathcal{N}_{s,0} + z_s\,\mathcal{N}_{s,1}
 + z_s^2\,\mathcal{N}_{s,2} + z_s^3\,\mathcal{N}_{s,3} ,
\label{eq:VLVL-Ns-expansion}
\end{align} 
$\mathcal{N}_{s,0}$ contains the unshifted terms obtained by removing hats in Eq.~\eqref{eq:VLVL-Ns-split}; the terms proportional to $\mathcal{N}_{s,1}$ cancel after summing over the two $z_{s,\pm}$ poles. Here,
\begin{align}
\mathcal{N}_{s,2}
&= -\frac{gf^{abc}}{m_1m_2}
\bigg\{
\frac{r_s^2}{2}
\Bigl[(p_1-p_2)+4\left(\eta_2-\eta_1\right)\Bigr]\cdot A_{(f_if_jV^c)}^{(\rho_3\rho_4)}
-\left(p_s\cdot r_s\right)(r_1-r_2)\cdot A_{(f_if_jV^c)}^{(\rho_3\rho_4)}
\bigg\} \nonumber \\
&\hspace{2.4em}+\frac{gf^{abc}\,g_{ffV}^{ijc}(T^c)_{ji}
        \left(m_1^2-m_2^2\right)\left(m_3-m_4\right)}
        {2m_I^2\,m_1m_2}\,r_s^2\,\overline v_4^{(\rho_4)}u_3^{(\rho_3)} ,
\label{eq:VLVL-Ns2}
\\[0.6em]
\mathcal{N}_{s,3}
&= \frac{gf^{abc}}{2m_1m_2}\,
r_s^2\,(r_1-r_2)\cdot   A_{(f_if_jV^c)}^{(\rho_3\rho_4)}.
\label{eq:VLVL-Ns3} 
\end{align}
where $r_s^2 = 2 r_1 \cdot r_2$, and $c$ denotes the species of the internal vector exchange. As we emphasized, in the massless limit defined in Eq.~\eqref{eq:massless_limit}, the unshifted numerator
$\mathcal{N}_{s,0}$ is at most $\mathcal{O}(\tau^0)$, while $\mathcal{N}_{s,2},\, \mathcal{N}_{s,3}$ are separately $\mathcal{O}(\tau^{-2})$, and the mass-singular terms are proportional to $z^2$ or above. Their divergence is a consequence of shifting $\hat \eta_i$. Also, note that the last term in  $\mathcal{N}_{s,2}$ is present due to the nonconservation of both currents when $m_1\neq m_2$ and $m_3\neq m_4$, and is also related to the $1/m_I^2$ term in $\hat \Pi^{\mu \nu}_L$. The rest of the terms in $\mathcal{N}_{s,2}, \, \mathcal{N}_{s,3}$ come from
\begin{align}
    -g^{\mu\nu} \hat A_{(V^aV^bV^c)\mu}^{(LL)}
   \hat A_{(f_if_jV^c)\nu}^{(\rho_3\rho_4)},
\end{align} 
which arises from the metric term in the transverse tensor and remains present as $m_I\to0$.
  After summing over the two $s$-channel residues, we find
\begin{align}
A_4^{(s)}
&= \frac{\mathcal{N}_{s,0}}{p_s^2-m_I^2}
\nonumber\\[0.3em]
&\quad
+ \frac{gf^{abc}\,g_{ffV}^{ijc}(T^c)_{ji}}{2m_1m_2}
  \Bigl[\left(p_1-p_2\right)+4\left(\eta_2-\eta_1\right)\Bigr]^{\nu}\,
  \overline v_4^{(\rho_4)}\gamma_{\nu}u_3^{(\rho_3)}
\nonumber\\[0.3em]
&\quad
+ \frac{gf^{abc}\,g_{ffV}^{ijc}(T^c)_{ji}
        \left(m_2^2-m_1^2\right)\left(m_3-m_4\right)}
       {2m_I^2\,m_1m_2}\,
  \overline v_4^{(\rho_4)}u_3^{(\rho_3)} .
\label{eq:VLVL-s-final}
\end{align}

 We have isolated the mass-singular contributions (the last two lines) in this channel as contact terms. If the four-point
amplitude has a smooth massless limit, we expect these contact terms to be canceled after including the factorized amplitudes from the $t,u$-channel fermion-exchanges or additional diagrams such as Higgs exchange.  This will be shown in the following.

We next consider the $t$-channel factorization through a massive fermion. We
define the momentum flowing through the internal fermion line as
\begin{align}
\hat p_t^{\mu}
&=
\hat p_1^{\mu}+\hat p_3^{\mu}
=
-\hat p_2^{\mu}-\hat p_4^{\mu}
=
p_t^{\mu}+z r_t^{\mu},
\qquad
r_t^{\mu}=r_1^{\mu}+r_3^{\mu} ,
\label{eq:t-channel-momentum}
\end{align}
with the pole condition $\hat p_t^2(z_t^\pm)=(m_F^t)^2$. The two
fermion--vector vertices carry the couplings $g_{ffV}^{ika}$ and
$g_{ffV}^{kjb}$ and the group factor
\begin{align}
(T^b)_{jk}(T^a)_{ki}=(T^bT^a)_{ji},
\end{align}
where $k$ labels the internal fermion of mass $m_F^t$. Using the three-point
amplitude in Eq.~\eqref{eq:LffV-expanded} rather than Eq.~\eqref{eq:Vff}, and
summing over the two factorization poles, the $t$-channel contribution becomes
\begin{align}
A_4^{(t)}
&= \frac{g_{ffV}^{ika}g_{ffV}^{kjb}\,(T^bT^a)_{ji}}{m_1m_2}
\Bigg[
\frac{
\overline v_4^{(\rho_4)}
\bigl(m_4-m_F^t-2\slashed\eta_2\bigr)
\bigl(\slashed p_t+m_F^t\bigr)
\bigl(m_F^t-m_3-2\slashed\eta_1\bigr)
u_3^{(\rho_3)}}
{p_t^2-\bigl(m_F^t\bigr)^2}
\nonumber\\[0.3em]
&\hspace{6em}
- \frac{1}{2}
  \Bigl[\left(p_1-p_2\right)+4\left(\eta_2-\eta_1\right)\Bigr]^{\nu}\,
  \overline v_4^{(\rho_4)}\gamma_{\nu}u_3^{(\rho_3)}
\nonumber\\[0.3em]
&\hspace{6em}
- \frac{1}{2}\left(2m_F^t-m_3-m_4\right)
  \overline v_4^{(\rho_4)}u_3^{(\rho_3)}
\Bigg].
\label{eq:VLVL-massive-t-final-group}
\end{align}
As in the $s$-channel case, the contact contributions that diverge in the massless limit originate from
shift-dependent parts of $\hat\eta_1^{\mu}$ and $\hat\eta_2^{\mu}$. In the massless limit the pole
term is $\mathcal{O}(\tau^0)$, the momentum-dependent part of the first contact term is $\mathcal{O}(\tau^{-2})$, and the last contact term is $\mathcal{O}(\tau^{-1})$. The $\eta$-dependent part of the first contact term is finite. The last contact term vanishes for the conserved current with $m_F^t=m_3=m_4$, while the first survives even for massless fermions.

The $u$-channel fermion exchange follows the same computation, with
$\hat p_u^{\mu}=\hat p_2^{\mu}+\hat p_3^{\mu}$, pole condition
$\hat p_u^2(z_u^\pm)=(m_F^u)^2$, and the two vertices exchanged,
\begin{align}
A_4^{(u)}
&= \frac{g_{ffV}^{ikb}g_{ffV}^{kja}\,(T^aT^b)_{ji}}{m_1m_2}
\Bigg[
\frac{
\overline v_4^{(\rho_4)}
\bigl(m_4-m_F^u-2\slashed\eta_1\bigr)
\bigl(\slashed p_u+m_F^u\bigr)
\bigl(m_F^u-m_3-2\slashed\eta_2\bigr)
u_3^{(\rho_3)}}
{p_u^2-\bigl(m_F^u\bigr)^2}
\nonumber\\[0.3em]
&\hspace{6em}
+ \frac{1}{2}
  \Bigl[\left(p_1-p_2\right)+4\left(\eta_2-\eta_1\right)\Bigr]^{\nu}\,
  \overline v_4^{(\rho_4)}\gamma_{\nu}u_3^{(\rho_3)}
\nonumber\\[0.3em]
&\hspace{6em}
- \frac{1}{2}\left(2m_F^u-m_3-m_4\right)
  \overline v_4^{(\rho_4)}u_3^{(\rho_3)}
\Bigg].
\label{eq:VLVL-massive-u-final-group}
\end{align}

The vector- and fermion-exchange contribution is given by
\begin{align}
    A_4 = A_4^{(s)}+ A_4^{(t)}+A_4^{(u)}.
    \label{eq:full}
\end{align}
We discuss the massless scaling and cancellation of the contact terms next.

\subsection{General Remarks}
\label{sec:general-remarks}
The $V_LV_Lff$ calculation illustrates how on-shell recursion separates
potential mass singularities from the pole contributions. In
Eqs.~\eqref{eq:VLVL-s-final}--\eqref{eq:VLVL-massive-u-final-group}, each
channel is written as a pole term that is finite in the simultaneous
massless limit of Eq.~\eqref{eq:massless_limit}, plus contact contributions. In this example, all terms scaling as
inverse powers of $\tau$ lie in contact contributions.

It is useful to separate the contact terms by their structure, since the
two classes cancel for different reasons. Writing
\begin{align}
\mathcal{X}
\equiv
\frac{1}{2m_1m_2}
\Bigl[\left(p_1-p_2\right)+4\left(\eta_2-\eta_1\right)\Bigr]^{\nu}\,
\overline v_4^{(\rho_4)}\gamma_{\nu}u_3^{(\rho_3)} ,
\qquad
\mathcal{Y} \equiv
\frac{1}{2m_1m_2}\,\overline v_4^{(\rho_4)}u_3^{(\rho_3)} ,
\end{align}
the three channels contribute
\begin{align}
A_4^{\rm contact}
&=
\Bigl[
gf^{abc}g_{ffV}^{ijc}(T^c)_{ji}
- g_{ffV}^{ika}g_{ffV}^{kjb}(T^bT^a)_{ji}
+ g_{ffV}^{ikb}g_{ffV}^{kja}(T^aT^b)_{ji}
\Bigr] \mathcal{X}
\nonumber\\
&\quad
+\Bigl[
\frac{gf^{abc}g_{ffV}^{ijc}(T^c)_{ji}
      \left(m_2^2-m_1^2\right)\left(m_3-m_4\right)}{m_I^2}
\nonumber\\
&
- g_{ffV}^{ika}g_{ffV}^{kjb}(T^bT^a)_{ji}\left(2m_F^t-m_3-m_4\right)
- g_{ffV}^{ikb}g_{ffV}^{kja}(T^aT^b)_{ji}\left(2m_F^u-m_3-m_4\right)
\Bigr]\mathcal{Y} .
\label{eq:contact-sum}
\end{align}
The first bracket can be canceled by imposing
\begin{equation}
    \begin{aligned}
&g_{ffV}^{ika}=g_{ffV}^{kjb}=g, \\&
[T^a,T^b]_{ij}=-f^{abc}(T^c)_{ij} .
\end{aligned}\label{eq:universality}
\end{equation}
The coupling universality in Eq.~\eqref{eq:universality} is understood to
apply to all fermion--vector vertices in the three channels. In the
simultaneous limit, the momentum-dependent part of $\mathcal{X}$ scales
as $\mathcal{O}(\tau^{-2})$, whereas its $\eta$-dependent part is
$\mathcal{O}(\tau^0)$ because $\eta_i=\mathcal{O}(\tau^2)$.
Equation~\eqref{eq:universality} cancels the entire coefficient of
$\mathcal{X}$, including the finite $\eta$-dependent contribution.
The cancellation of the leading divergence therefore imposes the same
coupling and representation structure as four-point consistency of the
massless theory~\cite{Benincasa:2007xk,McGady:2013sga}, without fixing the
broken mass spectrum.

The coefficient of $\mathcal{Y}$ in Eq.~\eqref{eq:contact-sum} is
$\mathcal{O}(\tau)$, while $\mathcal{Y}=\mathcal{O}(\tau^{-2})$.
This contribution is therefore $\mathcal{O}(\tau^{-1})$ and its
cancellation constrains the spectrum. The fermion-exchange terms vanish
for $m_F^t=m_F^u=m_3=m_4$, and the vector-exchange term vanishes for
$m_1=m_2$ or $m_3=m_4$. Alternatively, additional exchanges can provide
the required cancellation, as in a Higgs completion with appropriately
related scalar and Yukawa couplings.\footnote{For more discussion on
unitarization from the bottom up, see Ref.~\cite{Jeong:2026dzt}.}
Cancellation in this process alone does not establish consistency of the
same spectrum in every scattering process.

Writing $\epsilon_{L,i}^\mu=(\hat p_i^\mu-2\hat \eta_i^\mu)/m_i$ and using the
\textit{scalar-like} three-point amplitudes $A_{\phi^a\phi^bV^c}^{(\rho_3)}$ and
$A_{f_if_j\phi^c}^{(\rho_1\rho_2)}$, together with the remaining
$\eta$-dependent terms in Eqs.~\eqref{eq:LLV} and
\eqref{eq:LffV-expanded}, redistributes contributions between the pole
and contact parts relative to a calculation using the longitudinal
polarizations directly. Both constructions give the same channel
amplitudes. The advantage of the decomposed form is that it isolates the
mass-singular terms as contact contributions in each channel, leaving
pole terms that are finite in the simultaneous massless limit and making
the cancellation conditions explicit. Contact contributions can also contain terms that are finite in the
massless limit and need not cancel. For example, the four-vector contact
interaction in Eq.~\eqref{eq:VVVV-contact} is finite for
transversely polarized external vectors. 

To discuss the four-point current constraint, choose a longitudinal external vector as leg $1$ and write
\begin{align}
 A_{4,L}
 &=\epsilon_{L,1}^\mu A_{4\,\mu}
 =\frac{p_1^\mu}{m_1}A_{4\,\mu}
  -\frac{2\eta_1^\mu}{m_1}A_{4\,\mu}
 \nonumber\\
 &\equiv A_{\mathrm{scalar}}+A_{\eta}.
 \label{eq:4pt-longitudinal-decomposition}
\end{align}
Here $A_{4\,\mu}$ is the polarization-stripped current with the other
external states kept fixed. The \textit{scalar-like} component
$A_{\mathrm{scalar}}$ is obtained by replacing the polarization on leg
$1$ with $p_1^\mu/m_1$, while $A_{\eta}$ contains its
$-2\eta_1^\mu/m_1$ component. The four-point current constraint is
\begin{equation}
 p_1^\mu A_{4\,\mu}=\mathcal{O}(m_1),
 \label{eq:4pt-current-constraint}
\end{equation}
which is equivalent to requiring $A_{\mathrm{scalar}}$ to remain finite
in the simultaneous massless limit. In the \(V_LV_Lff\) calculation, all mass-singular contact contributions originate from the shift-dependent parts of \(\hat\eta_1\) and \(\hat\eta_2\). Their cancellation makes the physical longitudinal amplitude finite. To infer the current constraint in Eq.~\eqref{eq:4pt-current-constraint}, the scalar-like component must also be finite separately. In a Lagrangian formulation, the corresponding vector–Goldstone relation follows from the Slavnov–Taylor identities and applies with \(p_1\) off shell, provided the other external states are on shell and physical. Our calculation, however, uses on-shell external momenta. Deriving the corresponding off-shell current constraint directly from the construction presented here is left to future work.

Our decomposition parallels the five-component formulations of
Refs.~\cite{Wulzer:2013mza,Cuomo:2019siu,Chen:2022gxv} in separating the
longitudinal contribution into a scalar part and a residual vector part.
In those formulations, the Slavnov--Taylor identities replace the
$p^\mu/m$ contraction by a Goldstone amplitude, while the remaining
vector component is $\widetilde\epsilon_L^\mu=-2\eta^\mu/m$.
In our construction, 
after the mass-singular contact contributions cancel, the resulting
pole terms have the same scalar-plus-vector organization. The parallel
also applies to the internal line where contractions with $p_I^\mu/m_I$ are
expressed through scalar-like currents, leaving the transverse and
residual $\eta_I$-dependent contributions. 

\section{Conclusion}
\label{sec:conclusion}

In this work, we investigated how the massive on-shell amplitude formalism can guide off-shell decompositions. Using on-shell factorization and the ALT momentum shift, whose shift directions are proportional to transverse polarization vectors, we constructed an explicit decomposition of vector-boson exchange for the four-point amplitudes studied. The resulting transverse and longitudinal contributions can be expressed in terms of unshifted internal momenta, accounting for the physical degrees of freedom and making the massless limit transparent.

Our construction derives the internal-propagator decomposition from massive on-shell amplitudes through factorization, constructibility, and the three-point current constraint. Under the ALT shift, terms linear in $z$ in the full channel numerator cancel in the weighted sum over the two factorization roots, whereas higher-order terms can produce contact interactions. This gives an independent on-shell account of the structures appearing in the gauge-theory formulations of Refs.~\cite{Chanowitz:1985hj,Wulzer:2013mza,Chen:2016wkt,Chen:2019dkx,Cuomo:2019siu}. It also offers a constructive route to higher-point decompositions, for which the current constraint must be generalized beyond three points.

Furthermore, in the presence of external longitudinal vectors, the amplitude may exhibit mass singularities in individual channels. By a decomposition of the longitudinal polarization vector, the on-shell construction clearly identifies how these mass singularities propagate from lower-point to higher-point amplitudes. This also establishes an algorithm to isolate the singular parts as contact terms, and their cancellation in the full amplitude leaves well-behaved pole contributions in each channel. In the $VVff$ example, the cancellation of these contact contributions yields constraints on the couplings and mass spectrum directly from on-shell consistency. We leave the application to more general amplitudes for future work. For this purpose, it would be particularly useful to refine the large-$z$ analysis and find more general classes of constructible amplitudes, such as those in higher-derivative or higher-spin theories. The decomposition derived in this work will then help to understand the origin of the mass singularities, and eventually find UV completions to remove them.

\acknowledgments
We thank Kaoru Hagiwara for helpful discussions, in particular in introducing the FD gauge. 
ZL and IM are supported in part by the U.S. Department of Energy under Grant No.~DE-SC0011842 and by a Sloan Research Fellowship from the Alfred P. Sloan Foundation at the University of Minnesota. WK is supported in part by the European Union’s Horizon 2025 research and
innovation program under the Marie Sklodowska-Curie grant agreement No  101270027. We acknowledge support from the Simons Foundation Targeted Grant 920184 to the Fine Theoretical Physics Institute, enabling a visitor program that positively impacted our work.

We acknowledge assistance from Anthropic's Claude and OpenAI tools in preparing the manuscript. All calculations in this work were performed directly by the authors.

\appendix

\section{Spinor-helicity formalism and conventions}
\label{app:conventions}

\paragraph{Lorentz and spinor indices:}
We use the mostly-minus metric and fix the orientation by
\begin{equation}
 g_{\mu\nu}=\operatorname{diag}(1,-1,-1,-1),\qquad
 \varepsilon^{0123}=-1,\qquad \varepsilon_{0123}=+1.
 \label{eq:conv-metric}
\end{equation}
Symmetrization of two Lorentz indices includes a factor of
$1/2$, so that $2a^{(\mu}b^{\nu)}=a^\mu b^\nu+a^\nu b^\mu$.

Undotted and dotted spinor indices are denoted by $a,b$ and $\dot a,\dot b$,
respectively. The Pauli matrices enter through
\begin{equation}
 \sigma^\mu=(\mathbf 1,\vec\sigma),\qquad
 \bar\sigma^\mu=(\mathbf 1,-\vec\sigma),\qquad
 \sigma^\mu\bar\sigma^\nu+\sigma^\nu\bar\sigma^\mu
 =2g^{\mu\nu}\mathbf 1.
\end{equation}
The two matrix representations of a four-vector are
\begin{align}
 p_{a\dot a}&=p_\mu\sigma^\mu_{a\dot a}
 =\begin{pmatrix}p^0-p^3&-p^1+ip^2\\-p^1-ip^2&p^0+p^3\end{pmatrix},
 \label{eq:conv-pmatrix}\\
 p^{\dot a a}&=p_\mu\bar\sigma^{\mu\dot a a}
 =\begin{pmatrix}p^0+p^3&p^1-ip^2\\p^1+ip^2&p^0-p^3\end{pmatrix}.
\end{align}
In particular, $\det p_{a\dot a}=p^2$ and
$p^{\dot a a}q_{a\dot a}=2p\cdot q$.

For each two-component index space we use
\begin{equation}
 \epsilon^{12}=-\epsilon_{12}=1,\qquad
 \chi^a=\epsilon^{ab}\chi_b,\qquad
 \chi_a=\epsilon_{ab}\chi^b.
 \label{eq:conv-epsilon}
\end{equation}

\paragraph{Massless spinors:}
A null vector $k$ is represented by a rank-one matrix,
\begin{equation}
 k_{a\dot a}=|k\rangle_a[k|_{\dot a},\qquad
 k^{\dot a a}=|k]^{\dot a}\langle k|^a.
\end{equation}
Our bracket and mixed-contraction conventions are
\begin{align}
 \langle ij\rangle&=\langle i|^a|j\rangle_a,
 & [ij]&=[i|_{\dot a}|j]^{\dot a},\nonumber\\
 \langle i|\sigma^\mu|j]&=\langle i|^a\sigma^\mu_{a\dot a}|j]^{\dot a},
 &\langle i|p|j]&=\langle i|^a p_{a\dot a}|j]^{\dot a}.
 \label{eq:conv-brackets}
\end{align}

\paragraph{Massive momenta and the $\lambda+\eta$ decomposition:}
Our starting point for a massive external leg is the decomposition used in
the main text. For $p_i^2=m_i^2\ne0$, choose a null reference vector $n_i$
with $p_i\cdot n_i\ne0$ and define
\begin{equation}
 p_i^\mu=\lambda_i^\mu+\eta_i^\mu,\qquad
 \eta_i^\mu=\frac{m_i^2}{2p_i\cdot n_i}n_i^\mu,\qquad
 \lambda_i^\mu=p_i^\mu-\eta_i^\mu,\qquad
 \lambda_i^2=\eta_i^2=0.
 \label{eq:conv-null}
\end{equation}
Each null vector is represented by its own pair of spinors, giving
\begin{equation}
 (p_i)_{a\dot a}=|\lambda_i\rangle_a[\lambda_i|_{\dot a}
                    +|\eta_i\rangle_a[\eta_i|_{\dot a},\qquad
 2\lambda_i\cdot\eta_i=\langle\lambda_i\eta_i\rangle[\eta_i\lambda_i]=m_i^2.
 \label{eq:conv-plus}
\end{equation}
We fix the relative spinor normalization by
\begin{equation}
 \langle\lambda_i\eta_i\rangle=[\eta_i\lambda_i]=m_i,
 \qquad
 \langle\eta_i\lambda_i\rangle=[\lambda_i\eta_i]=-m_i.
 \label{eq:conv-null-normalization}
\end{equation}
The spinors $|\lambda_i\rangle$, $|\eta_i\rangle$, $|\lambda_i]$, and
$|\eta_i]$ are spinors of the two null constituents of $p_i$, and therefore
carry no massive little-group index. Rescaling $n_i$ leaves the decomposition
unchanged, while changing its direction changes the spin quantization axis.

\paragraph{Boldface massive spinors and little-group indices:}
The two null spinors can be assembled into massive spinors
$|\mathbf{i}\rangle^I$ and $[\mathbf{i}|^I$, with $I=1,2$.
We use boldface for the massive spinor label, including when its little-group
index is displayed explicitly; the constituent labels $\lambda_i$ and
$\eta_i$ remain unbolded. In our chosen basis,
\begin{align}
 |\mathbf{i}\rangle_a^I
 &=|\eta_i\rangle_a\delta^I_1+|\lambda_i\rangle_a\delta^I_2,
 \nonumber\\
 [\mathbf{i}|_{\dot a}^I
 &=[\lambda_i|_{\dot a}\delta^I_1-[\eta_i|_{\dot a}\delta^I_2.
 \label{eq:conv-basis}
\end{align}

The indices $I,J$ transform under the massive little group, which is
$SU(2)$ for real timelike momenta. We raise and lower them with
$\epsilon^{12}=-\epsilon_{12}=1$. Thus
$[\mathbf{i}|_{\dot a1}=[\eta_i|_{\dot a}$ and
$[\mathbf{i}|_{\dot a2}=[\lambda_i|_{\dot a}$, and the momentum becomes
\begin{equation}
 (p_i)_{a\dot a}=|\mathbf{i}\rangle_a^I[\mathbf{i}|_{\dot a I},\qquad
 (p_i)^{\dot a a}=|\mathbf{i}]_I^{\dot a}\langle\mathbf{i}|^{aI}.
 \label{eq:conv-massive}
\end{equation}
The normalization in Eq. \eqref{eq:conv-null-normalization} implies
\begin{equation}
 \det\bigl[|\mathbf{i}\rangle_a^I\bigr]
 =\det\bigl[[\mathbf{i}|_{\dot a I}\bigr]=m_i,
 \qquad
 \langle\mathbf{i}^I\mathbf{i}^J\rangle=-m_i\epsilon^{IJ},\qquad
 [\mathbf{i}_I\mathbf{i}_J]=-m_i\epsilon_{IJ}.
 \label{eq:conv-massive-normalization}
\end{equation}
For a spin-one external state the two indices associated with that leg
are symmetrized.

\paragraph{Dirac equations and wavefunctions in the $\lambda,\eta$ basis:}
The decomposition and normalization above give the two-component relations
\begin{align}
 (p_i)_{a\dot a}|\lambda_i]^{\dot a}
 &=m_i|\eta_i\rangle_a,
 & (p_i)_{a\dot a}|\eta_i]^{\dot a}
 &=-m_i|\lambda_i\rangle_a,\nonumber\\
 (p_i)^{\dot a a}|\eta_i\rangle_a
 &=m_i|\lambda_i]^{\dot a},
 & (p_i)^{\dot a a}|\lambda_i\rangle_a
 &=-m_i|\eta_i]^{\dot a}.
 \label{eq:conv-constituent-dirac}
\end{align}
Equivalently, the boldface massive spinors satisfy
\begin{equation}
 (p_i)_{a\dot a}|\mathbf{i}]^{\dot a I}
 =m_i|\mathbf{i}\rangle_a^I,\qquad
 (p_i)^{\dot a a}|\mathbf{i}\rangle_a^I
 =m_i|\mathbf{i}]^{\dot a I}.
 \label{eq:conv-massive-identities}
\end{equation}
We use the Weyl representation,
\begin{equation}
 \gamma^\mu=\begin{pmatrix}0&\sigma^\mu\\\bar\sigma^\mu&0\end{pmatrix},
 \qquad \gamma^5=\operatorname{diag}(-\mathbf1,\mathbf1),\qquad
 P_{L,R}=\frac{1\mp\gamma^5}{2},\qquad
 \slashed p=p_\mu\gamma^\mu.
\end{equation}
Suppressing Lorentz spinor indices, the particle and antiparticle
wavefunctions in the chosen spin basis are
\begin{align}
 u^1(p_i)&=\begin{pmatrix}|\eta_i\rangle\\|\lambda_i]\end{pmatrix},
 &u^2(p_i)&=\begin{pmatrix}|\lambda_i\rangle\\-|\eta_i]\end{pmatrix},\nonumber\\
 v^1(p_i)&=\begin{pmatrix}|\eta_i\rangle\\-|\lambda_i]\end{pmatrix},
 &v^2(p_i)&=\begin{pmatrix}|\lambda_i\rangle\\|\eta_i]\end{pmatrix}.
 \label{eq:conv-wavefunctions}
\end{align}
Eq. \eqref{eq:conv-constituent-dirac} directly gives
\begin{equation}
 (\slashed p_i-m_i)u^I(p_i)=0,\qquad
 (\slashed p_i+m_i)v^I(p_i)=0.
 \label{eq:conv-dirac}
\end{equation}
Similarly,
\begin{equation}
 \bar u^I(p_i)=\bigl(-\langle\mathbf{i}|^{aI},[\mathbf{i}|_{\dot a}^{I}\bigr),
 \qquad
 \bar v^I(p_i)=\bigl(\langle\mathbf{i}|^{aI},[\mathbf{i}|_{\dot a}^{I}\bigr).
 \label{eq:conv-duals}
\end{equation}
They obey
$\bar u^I(\slashed p_i-m_i)=0$ and $\bar v^I(\slashed p_i+m_i)=0$, with spin sums
\begin{equation}
 \epsilon_{IJ}u^I\bar u^J=\slashed p_i+m_i,\qquad
 \epsilon_{IJ}v^I\bar v^J=\slashed p_i-m_i.
 \label{eq:conv-fermion-sums}
\end{equation}

For momentum reversal we keep the angle spinors fixed and reverse the
sign of both square spinors:
\begin{equation}
 |\lambda_i\rangle\longmapsto|\lambda_i\rangle,\quad
 |\eta_i\rangle\longmapsto|\eta_i\rangle,\qquad
 [\lambda_i|\longmapsto-[\lambda_i|,\quad
 [\eta_i|\longmapsto-[\eta_i|.
\end{equation}
Both null momenta then reverse sign, so $p_i\mapsto-p_i$. In boldface
notation this is
\begin{equation}
 |-\mathbf{i}\rangle^I=|\mathbf{i}\rangle^I,\qquad
 |-\mathbf{i}]^I=-|\mathbf{i}]^I,\qquad
 u^I(-p_i)=v^I(p_i),\qquad v^I(-p_i)=u^I(p_i).
 \label{eq:conv-crossing}
\end{equation}
\paragraph{Explicit helicity basis:}
For a future-directed momentum $p^\mu=(E,\boldsymbol p)$, let
$k=|\boldsymbol p|$, $c=\cos(\theta/2)$, and
$s=e^{i\varphi}\sin(\theta/2)$. The null constituent spinors may be chosen as
\begin{align}
 |\lambda\rangle&=\sqrt{E+k}\begin{pmatrix}-s^*\\c\end{pmatrix},
 &[\lambda|&=\sqrt{E+k}\begin{pmatrix}-s&c\end{pmatrix},\nonumber\\
 |\eta\rangle&=\sqrt{E-k}\begin{pmatrix}c\\s\end{pmatrix},
 &[\eta|&=\sqrt{E-k}\begin{pmatrix}c&s^*\end{pmatrix}.
 \label{eq:conv-explicit-spinors}
\end{align}
This realizes $n\propto(1,-\boldsymbol p/k)$ for $k\ne0$, with
\begin{equation}
 \lambda^\mu=\frac{E+k}{2}(1,\boldsymbol p/k),\qquad
 \eta^\mu=\frac{E-k}{2}(1,-\boldsymbol p/k).
\end{equation}
In this basis $u^1$ and $u^2$ have helicities $+1/2$ and $-1/2$,
respectively. For a general reference direction, their indices instead
label the two spin projections in the basis selected by $n$.
At $E\gg m$, the $\lambda$ spinors scale as $O(\sqrt E)$ and the
$\eta$ spinors as $O(m/\sqrt E)$, since $E-k=m^2/(E+k)$.

\paragraph{Vector polarizations and completeness:}
As discussed in the main text, with the contractions in \eqref{eq:conv-brackets}, the polarization vectors are
\begin{equation}
 \epsilon_+^\mu=\frac{\langle\eta|\sigma^\mu|\lambda]}{\sqrt2\langle\eta\lambda\rangle},
 \qquad
 \epsilon_-^\mu=\frac{\langle\lambda|\sigma^\mu|\eta]}{\sqrt2[\lambda\eta]},
 \qquad
 \epsilon_L^\mu=\frac{\lambda^\mu-\eta^\mu}{m}
              =\frac{p^\mu}{m}-\frac{m n^\mu}{p\cdot n}.
 \label{eq:conv-polarizations}
\end{equation}
For real momenta the phases can be chosen so that
$\epsilon_-=(\epsilon_+)^*$ and $\epsilon_L$ is real. Then
\begin{equation}
 p\cdot\epsilon_\rho=0,\qquad
 \epsilon_\rho\cdot\epsilon_{\rho'}^*=-\delta_{\rho\rho'},
 \qquad \rho,\rho'=\pm,L.
\end{equation}

The polarization tensors $\Pi_\rho^{\mu\nu}
=\epsilon_\rho^\mu\epsilon_{\bar\rho}^\nu$ are
\begin{align}
 \Pi_\pm^{\mu\nu}
 &=\frac12\left(-g^{\mu\nu}
    +\frac{p^\mu n^\nu+p^\nu n^\mu}{p\cdot n}
    -\frac{m^2n^\mu n^\nu}{(p\cdot n)^2}\right)
   \mathbin{\pm}\frac{i\varepsilon^{\mu\nu\alpha\beta}
                   p_\alpha n_\beta}{2p\cdot n},
 \label{eq:conv-projectors}\\
 \Pi_L^{\mu\nu}
 &=\frac{p^\mu p^\nu}{m^2}
   -\frac{p^\mu n^\nu+p^\nu n^\mu}{p\cdot n}
   +\frac{m^2n^\mu n^\nu}{(p\cdot n)^2}.
\end{align}
Writing $\Pi_T=\Pi_++\Pi_-$, we have
\begin{equation}
 \Pi_T^{\mu\nu}+\Pi_L^{\mu\nu}
 =-g^{\mu\nu}+\frac{p^\mu p^\nu}{m^2},\qquad
 (\Pi_\rho)^\mu{}_{\alpha}(\Pi_{\rho'})^{\alpha\nu}
 =-\delta_{\rho\rho'}\Pi_\rho^{\mu\nu}.
 \label{eq:conv-completeness}
\end{equation}

For a massless vector, the transverse polarizations follow by replacing
$\lambda$ with its momentum $k$ and $\eta$ with any non-collinear null
reference vector $n$ in the first two expressions of
Eq. \eqref{eq:conv-polarizations}. Their sum is
\begin{equation}
 \sum_{\rho=\pm}\epsilon_\rho^\mu(k;n)\epsilon_{\bar\rho}^\nu(k;n)
 =-g^{\mu\nu}+\frac{k^\mu n^\nu+k^\nu n^\mu}{k\cdot n}.
\end{equation}
Reference-vector dependence drops out after contraction with conserved
currents.

\section{All-line transverse shift}
\label{app:alt}

\subsection{Definition and external wavefunctions}
\label{app:alt-definition}
For an $n$-point amplitude with all external momenta incoming, we introduce
\begin{equation}
 \hat p_i^\mu(z)=p_i^\mu+zr_i^\mu,\qquad
 r_i^2=p_i\cdot r_i=0,\qquad \sum_{i=1}^n r_i^\mu=0.
 \label{eq:alt-shift}
\end{equation}
The external mass shells and momentum conservation are then preserved for
every complex $z$. In the ALT shift,
\begin{equation}
 r_i^\mu=c_i\epsilon_{\sigma_i}^\mu(p_i),\qquad
 \sigma_i\in\{+,-\},\qquad [z]=0,\qquad[c_i]=1.
 \label{eq:alt-direction}
\end{equation}
The transverse direction is chosen after specifying the external state, so
that its wavefunction is independent of $z$. For a transverse vector it is
the direction of that vector's polarization. For a longitudinal vector
either transverse direction can be used; shifting along $\epsilon_L$ would
not preserve the mass shell because $\epsilon_L^2=-1$.
For a scalar there is no polarization wavefunction to constrain the choice.

In the helicity basis of Appendix~\ref{app:conventions}, all transverse
polarizations have zero temporal component in a common frame. Momentum
conservation therefore gives three homogeneous linear conditions on the
$c_i$, admitting a nontrivial solution for $n\geq4$.

To display the spinor shifts, introduce a dimensionless coefficient $b_i$.
For a positive transverse vector, or the fermion wavefunctions $u^1,v^1$
in Eq. \eqref{eq:conv-wavefunctions}, one possible deformation is
\begin{equation}
 |\hat\lambda_i\rangle
 =|\lambda_i\rangle+zb_i|\eta_i\rangle,
 \qquad (r_i)_{a\dot a}=b_i|\eta_i\rangle_a[\lambda_i|_{\dot a},
 \label{eq:alt-positive}
\end{equation}
with the other three constituent spinors fixed. For a negative transverse
vector, or $u^2,v^2$, we may instead choose
\begin{equation}
 [\hat\lambda_i|
 =[\lambda_i|+zb_i[\eta_i|,
 \qquad (r_i)_{a\dot a}=b_i|\lambda_i\rangle_a[\eta_i|_{\dot a}.
 \label{eq:alt-negative}
\end{equation}
Using Eq. \eqref{eq:conv-polarizations}, both choices agree with
Eq. \eqref{eq:alt-direction} when
\begin{equation}
 c_i=-\frac{m_i b_i}{\sqrt2}.
 \label{eq:alt-normalization}
\end{equation}

For a longitudinal external vector, we shift both null constituents equally:
\begin{equation}
 \begin{cases}
 |\hat\lambda_i\rangle=|\lambda_i\rangle+\dfrac{zb_i}{2}|\eta_i\rangle,\\[3pt]
 [\hat\eta_i|=[\eta_i|+\dfrac{zb_i}{2}[\lambda_i|,
 \end{cases}
 \quad\text{or}\quad
 \begin{cases}
 [\hat\lambda_i|=[\lambda_i|+\dfrac{zb_i}{2}[\eta_i|,\\[3pt]
 |\hat\eta_i\rangle=|\eta_i\rangle+\dfrac{zb_i}{2}|\lambda_i\rangle.
 \end{cases}
 \label{eq:alt-longitudinal-spinors}
\end{equation}
The remaining spinors are fixed. The two shifts act along the positive and
negative transverse directions, respectively, with $r_i$ as in
Eq. \eqref{eq:alt-positive} or Eq. \eqref{eq:alt-negative}. In either case,
\begin{equation}
 \hat\lambda_i^\mu=\lambda_i^\mu+\frac z2r_i^\mu,\qquad
 \hat\eta_i^\mu=\eta_i^\mu+\frac z2r_i^\mu,\qquad
 \hat\epsilon_{L,i}^\mu
 =\frac{\hat\lambda_i^\mu-\hat\eta_i^\mu}{m_i}
 =\epsilon_{L,i}^\mu.
 \label{eq:alt-longitudinal-fixed}
\end{equation}
 Although the longitudinal polarization is fixed,
its two pieces $\hat p_i/m_i$ and $-2\hat\eta_i/m_i$ are shifted.
This is the origin of the separate shift-dependent terms studied in the
longitudinal-amplitude calculation.

For a massless leg, write $p_i=|i\rangle[i|$ and shift its momentum by a null reference spinor. A positive-helicity state permits a shift
of $|i\rangle$ along the reference angle spinor, and a negative-helicity
state permits a shift of $[i|$ along the reference square spinor. 

\subsection{Factorization}
\label{app:alt-recursion}
For a channel $I$, define the incoming momentum on the left subamplitude
and its virtuality by
\begin{equation}
 \hat p_I(z)=p_I+zr_I,\qquad
 p_I=-\sum_{i\in L}p_i,\qquad r_I=-\sum_{i\in L}r_i,\qquad
 \Delta_I\equiv p_I^2-m_I^2.
\end{equation}
Although each $r_i$ is null, a generic sum $r_I$ is not. The factorization
condition is
\begin{equation}
 D_I(z)\equiv\hat p_I^2(z)-m_I^2
 =\Delta_I+2z\,p_I\cdot r_I+z^2r_I^2=0.
 \label{eq:alt-pole-polynomial}
\end{equation}
For $r_I^2\ne0$ it has two roots,
\begin{gather}
 z_I^\pm=\frac{-p_I\cdot r_I
 \pm\sqrt{(p_I\cdot r_I)^2-r_I^2\Delta_I}}{r_I^2},\nonumber\\
 z_I^++z_I^-=-\frac{2p_I\cdot r_I}{r_I^2},\qquad
 z_I^+z_I^-=\frac{\Delta_I}{r_I^2}.
 \label{eq:alt-roots}
\end{gather}
We initially take the roots to be distinct and the unshifted kinematics to
be away from a physical pole. Coincident roots are handled by a limit of
the combined channel expression.

Let $N_I(z_I)$ denote the product of the two on-shell subamplitudes summed over the physical intermediate states, with the right subamplitude carrying
momentum $-\hat p_I$. 
The recursion relation gives the $n$-point amplitude $A_n$
\begin{equation}
 A_n=-\sum_{I,\,z_I}\operatorname*{Res}_{z=z_I}
       \frac{N_I(z)}{zD_I(z)}+B_\infty,\qquad
 B_\infty=-\operatorname*{Res}_{z=\infty}\frac{\hat A_n(z)}{z}.
 \label{eq:alt-residues}
\end{equation}
For channels with two factorization roots, evaluating the residues gives
\begin{equation}
 A_n=\sum_I\frac{1}{\Delta_I}
 \frac{z_I^+N_I(z_I^-)-z_I^-N_I(z_I^+)}{z_I^+-z_I^-}
 +B_\infty.
 \label{eq:alt-two-root}
\end{equation}
If $r_I^2=0$ and $p_I\cdot r_I\ne0$, the channel instead has one root,
\begin{equation}
 z_I=-\frac{\Delta_I}{2p_I\cdot r_I},\qquad
 A_{n,I}=\frac{N_I(z_I)}{\Delta_I}.
 \label{eq:alt-one-root}
\end{equation}

The two-root combination explains why terms linear in $z$ drop out of
the channel sum. For a polynomial numerator of degree at most three,
\begin{align}
 N_I(z)&=N_{I,0}+zN_{I,1}+z^2N_{I,2}+z^3N_{I,3},\nonumber\\
 \frac{z_I^+N_I(z_I^-)-z_I^-N_I(z_I^+)}{z_I^+-z_I^-}
 &=N_{I,0}-z_I^+z_I^-N_{I,2}
   -z_I^+z_I^-(z_I^++z_I^-)N_{I,3}.
   \label{eq:sum_poles}
\end{align}
Hence its channel contribution is
\begin{equation}
 A_{n,I}=\frac{N_{I,0}}{\Delta_I}
 -\frac{N_{I,2}}{r_I^2}
 +\frac{2p_I\cdot r_I}{(r_I^2)^2}N_{I,3}.
 \label{eq:alt-contact}
\end{equation}
The last two terms carry no pole in $\Delta_I$. They are the source of
the contact contributions isolated in the four-point examples. The absence
of this channel pole does not by itself establish locality in all channels
or independence of the shift parameters; those properties must be checked
in the resulting amplitude.

\subsection{\texorpdfstring{Large-$z$ behavior and constructibility}{Large-z behavior and constructibility}}
\label{app:alt-large-z}
The condition $\hat A_n(z)\to0$ as $|z|\to\infty$ is sufficient for
$B_\infty=0$. For an $n$-point amplitude $A_n$ with coupling product $g_{\mathrm{prod}}$, strip the
external wavefunctions schematically as
\begin{equation}
 A_n=g_{\mathrm{prod}}F_n\prod_{\text{vectors}}\epsilon_i
                        \prod_{\text{fermions}}w_j,\qquad
 [F_n]=4-n-d_g-\frac{N_F}{2},\qquad d_g=[g_{\mathrm{prod}}].
 \label{eq:alt-dimensions}
\end{equation}
Here $w_j$ denotes any external Dirac wavefunction, and $N_F$
counts external fermions. The amplitude has mass dimension $4-n$, each
fermion wavefunction has dimension $1/2$, and each vector polarization
has dimension zero. Coupling structures with different $d_g$ are counted
separately, with the weakest bound controlling their sum.

Under a generic ALT shift, internal bosonic denominators scale as $z^2$,
and a fermion propagator scales as $z^{-1}$ after including its numerator.
The external wavefunctions are fixed. Once inverse-mass enhancements from
internal longitudinal propagation are controlled, the largest power of
the shifted momenta in $F_n$ is bounded by its mass dimension. This gives
the estimate of the EW ALT-shift analysis,
\begin{equation}
 \hat A_n(z)=O(z^\gamma),\qquad
 \gamma\leq4-n-d_g-\frac{N_F}{2}.
 \label{eq:alt-dimension-bound}
\end{equation}

A massive-vector
numerator contains $\hat p^\mu\hat p^\nu/m^2$, which would
invalidate this counting if its two powers of shifted momentum were left
uncompensated. We assume that current constraints, Ward
identities, or mass factors in dimensionful couplings remove that
enhancement. At three points the needed constraints are
those of the minimal vertices in the main text. 

There is a further improvement when an external vector is transverse.
For such a leg, say leg 1, $r_1=c_1\epsilon_1$, and hence
$\hat p_1/z\to r_1\propto\epsilon_1$. The leading term that saturates
Eq. \eqref{eq:alt-dimension-bound} has the schematic form
\begin{align}
 \hat A_n(z)\big|_{\mathrm{leading}}
 &=g_{\mathrm{prod}}z^\kappa\epsilon_1^\mu
 F^{(0)}_\mu(r_1,\ldots,r_n)\nonumber\\
 &=\frac{g_{\mathrm{prod}}}{c_1}z^\kappa r_1^\mu
 F^{(0)}_\mu(r_1,\ldots,r_n), \qquad \kappa=4-n-d_g-\frac{N_F}{2}.
\end{align}
Assuming the
Ward identity of the massless gauge theory, the last contraction vanishes. It removes at least one power of $z$:
\begin{equation}
 \gamma\leq4-n-d_g-\frac{N_F}{2}-c_T,\qquad
 c_T=\begin{cases}
 1,&\text{at least one external transverse gauge boson},\\
 0,&\text{otherwise}.
 \end{cases}
 \label{eq:alt-ward-bound}
\end{equation}

For renormalizable interactions, $d_g\geq0$. Under the assumptions above,
Eq. \eqref{eq:alt-ward-bound} therefore ensures constructibility for $n\geq5$,
and at four points whenever $N_F\geq2$ or $c_T=1$. In particular, the
$ffff$ and $VVff$ amplitudes considered here have vanishing boundary terms,
including $V_LV_Lff$. Four-vector amplitudes with at least one transverse
leg also have $B_\infty=0$.


\section{Null decomposition in a factorization channel}
\label{app:null-decomposition}

The separation of an internal massive-vector contribution into transverse
and longitudinal parts requires a choice of spin basis at the factorization
pole. To express the result in terms of unshifted momenta, we choose a
reference vector that is fixed after defining the complex deformation. In this
appendix we specify details of such a choice and distinguish the on-shell null
decomposition from its off-shell continuation. Throughout,
$m_I\ne0$ is the physical mass of the exchanged vector and is independent of the virtuality of the propagator.

\subsection{Existence of a fixed null reference vector}
\label{app:null-existence}
For a channel with
\begin{equation}
 \hat p_I^\mu(z)=p_I^\mu+zr_I^\mu,\qquad
 D_I(z)=\hat p_I^2(z)-m_I^2,
\end{equation}
we demand that the reference vector satisfy all three conditions
\begin{equation}
 n_I^2=0,\qquad r_I\cdot n_I=0,\qquad
 d_I\equiv p_I\cdot n_I\ne0.
 \label{eq:null-reference-conditions}
\end{equation}
The first condition permits a null decomposition at the massive pole.
The second makes its denominator independent of the shift,
\begin{equation}
 \hat p_I(z)\cdot n_I=p_I\cdot n_I=d_I,
 \label{eq:null-fixed-denominator}
\end{equation}
and the third ensures that the decomposition is well defined. Here $n_I$ may depend on the
unshifted kinematics and on the chosen shift, but is held fixed as $z$
varies. The same $n_I$ is used at both roots of $D_I(z)$, but different channels
may use different reference vectors.

For $r_I\ne0$, a solution of Eq. \eqref{eq:null-reference-conditions} exists
over complexified four-dimensional momentum space if and only if
\begin{equation}
 p_I\notin\operatorname{span}\{r_I\}=\{\alpha r_I:\alpha\in\mathbb{C}\}.
 \label{eq:null-existence-condition}
\end{equation}
If $p_I=\alpha r_I$, every vector orthogonal to
$r_I$ is also orthogonal to $p_I$. To see sufficiency, consider the
three-dimensional space $r_I^\perp$ which can be spanned by three \textit{complex} null vectors. If $p_I$ were
orthogonal to every such null vector, it would belong to
$(r_I^\perp)^\perp=\operatorname{span}\{r_I\}$, contrary to Eq.
\eqref{eq:null-existence-condition}.

A choice when $r_I^2\ne0$ starts with an auxiliary null vector
$t$ satisfying $r_I\cdot t\ne0$ and writing
\begin{equation}
 r_I=k+\ell,\qquad
 \ell^\mu=\frac{r_I^2}{2r_I\cdot t}t^\mu,\qquad
 k^\mu=r_I^\mu-\ell^\mu,\qquad k^2=\ell^2=0.
\end{equation}
The spinor products
\begin{equation}
 (n_I^{(1)})_{a\dot a}=|k\rangle_a[\ell|_{\dot a},\qquad
 (n_I^{(2)})_{a\dot a}=|\ell\rangle_a[k|_{\dot a}
 \label{eq:null-cross-spinors}
\end{equation}
are null and orthogonal to both $k$ and $\ell$, hence to $r_I$.
Their overall normalization is immaterial. One must still check
$p_I\cdot n_I\ne0$. If a candidate fails this check, one changes the
candidate or the auxiliary decomposition of $r_I$.

For $r_I^2=0$, $r_I\ne0$, the simple choice $n_I=r_I$ works whenever
$p_I\cdot r_I\ne0$. If $p_I\cdot r_I=0$ but $p_I$ is not proportional
to $r_I$, one can instead use a vector from either of the families
\begin{equation}
 (n_I)_{a\dot a}=|r_I\rangle_a[\xi|_{\dot a},
 \qquad\text{or}\qquad
 (n_I)_{a\dot a}=|\xi\rangle_a[r_I|_{\dot a},
\end{equation}
choosing the reference spinor so that $d_I\ne0$.

\subsection{Null decomposition away from the pole}
\label{app:null-off-shell}
When Eq. \eqref{eq:null-reference-conditions} is satisfied, define
\begin{equation}
 \eta_I^\mu=\frac{m_I^2}{2d_I}n_I^\mu,\qquad
 \lambda_I^\mu=p_I^\mu-\eta_I^\mu,\qquad
 \hat\lambda_I^\mu(z)=\lambda_I^\mu+zr_I^\mu.
 \label{eq:null-fixed-mass-split}
\end{equation}
Then
\begin{equation}
 \hat p_I(z)=\hat\lambda_I(z)+\eta_I,\qquad
 \eta_I^2=0,\qquad
 \hat\lambda_I(z)\cdot n_I=d_I,\qquad
 2\hat\lambda_I(z)\cdot\eta_I=m_I^2.
\end{equation}
However, the other constituent satisfies
\begin{equation}
 \hat\lambda_I^2(z)=D_I(z),\qquad
 \lambda_I^2=p_I^2-m_I^2.
 \label{eq:null-virtuality}
\end{equation}
The decomposition is therefore a sum of two null momenta only at the
factorization poles $z=z_I^\pm$, where $D_I(z_I^\pm)=0$.
At those poles one may use
\begin{equation}
 [\hat p_I(z_I^\pm)]_{a\dot a}
 =|\hat\lambda_I(z_I^\pm)\rangle_a
   [\hat\lambda_I(z_I^\pm)|_{\dot a}
  +|\eta_I\rangle_a[\eta_I|_{\dot a}.
\end{equation}
For generic $z$, $\hat\lambda_I$ is not rank one and cannot be replaced
by a single angle--square spinor product. Away from the pole,
\eqref{eq:null-fixed-mass-split} is a vector decomposition defining a
continuation of the pole data. It should not be interpreted as an off-shell
massive spinor-helicity factorization with unchanged on-shell spinors.

\subsection{Continuation of the polarization tensors}
\label{app:null-polynomial}
At a massive pole the transverse sum and the longitudinal contribution
are the usual on-shell tensors. We continue their vector expressions as
\begin{align}
 \hat\Pi_{I,T}^{\mu\nu}(z)
 &=-g^{\mu\nu}
   +\frac{\hat p_I^\mu n_I^\nu+\hat p_I^\nu n_I^\mu}{d_I}
   -\frac{m_I^2n_I^\mu n_I^\nu}{d_I^2},\nonumber\\
 \hat\Pi_{I,L}^{\mu\nu}(z)
 &=\frac{\hat p_I^\mu\hat p_I^\nu}{m_I^2}
   -\frac{\hat p_I^\mu n_I^\nu+\hat p_I^\nu n_I^\mu}{d_I}
   +\frac{m_I^2n_I^\mu n_I^\nu}{d_I^2}.
 \label{eq:null-continued-tensors}
\end{align}
Here and below $\hat p_I=\hat p_I(z)$. These expressions have
the polynomial dependence used in the main text:
\begin{align}
 \hat\Pi_{I,T}^{\mu\nu}(z)
 &=\Pi_{I,T,0}^{\mu\nu}
   +z\frac{2r_I^{(\mu}n_I^{\nu)}}{d_I},\nonumber\\
 \hat\Pi_{I,L}^{\mu\nu}(z)
 &=\Pi_{I,L,0}^{\mu\nu}
   +z\left(\frac{2p_I^{(\mu}r_I^{\nu)}}{m_I^2}
           -\frac{2r_I^{(\mu}n_I^{\nu)}}{d_I}\right)
   +z^2\frac{r_I^\mu r_I^\nu}{m_I^2}.
 \label{eq:null-tensor-expansion}
\end{align}
The constant tensors are obtained by setting $z=0$ in
\eqref{eq:null-continued-tensors}:
\begin{align}
 \Pi_{I,T,0}^{\mu\nu}
 &=-g^{\mu\nu}
   +\frac{p_I^\mu n_I^\nu+p_I^\nu n_I^\mu}{p_I\cdot n_I}
   -\frac{m_I^2 n_I^\mu n_I^\nu}{(p_I\cdot n_I)^2},\\
 \Pi_{I,L,0}^{\mu\nu}
 &=\frac{p_I^\mu p_I^\nu}{m_I^2}
   -\frac{p_I^\mu n_I^\nu+p_I^\nu n_I^\mu}{p_I\cdot n_I}
   +\frac{m_I^2 n_I^\mu n_I^\nu}{(p_I\cdot n_I)^2}.
\end{align}
The coefficient is the physical mass squared $m_I^2$, including away from
the pole; replacing it by $p_I^2$ would change the off-shell continuation.
Because $d_I$ is nonzero and independent
of $z$, the reference choice introduces no additional finite-$z$ poles in
these tensors. The tensors in \eqref{eq:null-continued-tensors} are physical polarization
projectors only on the massive pole. For example, define
\begin{equation}
 \ell_I^\mu(z)=\frac{\hat p_I^\mu(z)}{m_I}-\frac{m_I}{d_I}n_I^\mu,
 \qquad \hat\Pi_{I,L}^{\mu\nu}=\ell_I^\mu\ell_I^\nu.
\end{equation}
Direct contraction gives
\begin{equation}
 \hat p_I\cdot\ell_I=\frac{D_I(z)}{m_I},\qquad
 \ell_I^2=-1+\frac{D_I(z)}{m_I^2},\qquad
 \hat p_{I\mu}\hat\Pi_{I,T}^{\mu\nu}
 =\frac{D_I(z)}{d_I}n_I^\nu.
 \label{eq:null-off-shell-tests}
\end{equation}
Thus the continued longitudinal vector is not transverse or unit-normalized
away from the pole, and the continued transverse tensor is not transverse
to the off-shell momentum. Nevertheless their sum is exactly
\begin{equation}
 \hat\Pi_{I,T}^{\mu\nu}(z)+\hat\Pi_{I,L}^{\mu\nu}(z)
 =-g^{\mu\nu}+\frac{\hat p_I^\mu(z)\hat p_I^\nu(z)}{m_I^2}
 \label{eq:null-total}
\end{equation}
for all $z$. This identity explains why the construction reproduces the
massive-vector numerator while its individual continued pieces need not
have the properties of on-shell projectors.

\bibliographystyle{JHEP}
\bibliography{ref}

\providecommand{\href}[2]{#2}\begingroup\raggedright\begin{thebibliography}{10}

\bibitem{Britto:2004ap}
R.~Britto, F.~Cachazo and B.~Feng, \emph{{New recursion relations for tree amplitudes of gluons}}, \href{https://doi.org/10.1016/j.nuclphysb.2005.02.030}{\emph{Nucl. Phys. B} {\bfseries 715} (2005) 499} [\href{https://arxiv.org/abs/hep-th/0412308}{{\ttfamily hep-th/0412308}}].

\bibitem{Britto:2005fq}
R.~Britto, F.~Cachazo, B.~Feng and E.~Witten, \emph{{Direct proof of tree-level recursion relation in Yang-Mills theory}}, \href{https://doi.org/10.1103/PhysRevLett.94.181602}{\emph{Phys. Rev. Lett.} {\bfseries 94} (2005) 181602} [\href{https://arxiv.org/abs/hep-th/0501052}{{\ttfamily hep-th/0501052}}].

\bibitem{Bedford:2005yy}
J.~Bedford, A.~Brandhuber, B.J.~Spence and G.~Travaglini, \emph{{A Recursion relation for gravity amplitudes}}, \href{https://doi.org/10.1016/j.nuclphysb.2005.016}{\emph{Nucl. Phys. B} {\bfseries 721} (2005) 98} [\href{https://arxiv.org/abs/hep-th/0502146}{{\ttfamily hep-th/0502146}}].

\bibitem{Cachazo:2005ca}
F.~Cachazo and P.~Svrcek, \emph{{Tree level recursion relations in general relativity}},  \href{https://arxiv.org/abs/hep-th/0502160}{{\ttfamily hep-th/0502160}}.

\bibitem{Badger:2005zh}
S.D.~Badger, E.W.N.~Glover, V.V.~Khoze and P.~Svrcek, \emph{{Recursion relations for gauge theory amplitudes with massive particles}}, \href{https://doi.org/10.1088/1126-6708/2005/07/025}{\emph{JHEP} {\bfseries 07} (2005) 025} [\href{https://arxiv.org/abs/hep-th/0504159}{{\ttfamily hep-th/0504159}}].

\bibitem{Badger:2005jv}
S.D.~Badger, E.W.N.~Glover and V.V.~Khoze, \emph{{Recursion relations for gauge theory amplitudes with massive vector bosons and fermions}}, \href{https://doi.org/10.1088/1126-6708/2006/01/066}{\emph{JHEP} {\bfseries 01} (2006) 066} [\href{https://arxiv.org/abs/hep-th/0507161}{{\ttfamily hep-th/0507161}}].

\bibitem{Ozeren:2006ft}
K.J.~Ozeren and W.J.~Stirling, \emph{{Scattering amplitudes with massive fermions using BCFW recursion}}, \href{https://doi.org/10.1140/epjc/s10052-006-0007-7}{\emph{Eur. Phys. J. C} {\bfseries 48} (2006) 159} [\href{https://arxiv.org/abs/hep-ph/0603071}{{\ttfamily hep-ph/0603071}}].

\bibitem{Boels:2007pj}
R.~Boels and C.~Schwinn, \emph{{CSW rules for a massive scalar}}, \href{https://doi.org/10.1016/j.physletb.2008.02.038}{\emph{Phys. Lett. B} {\bfseries 662} (2008) 80} [\href{https://arxiv.org/abs/0712.3409}{{\ttfamily 0712.3409}}].

\bibitem{Arkani-Hamed:2008owk}
N.~Arkani-Hamed, F.~Cachazo and J.~Kaplan, \emph{{What is the Simplest Quantum Field Theory?}}, \href{https://doi.org/10.1007/JHEP09(2010)016}{\emph{JHEP} {\bfseries 09} (2010) 016} [\href{https://arxiv.org/abs/0808.1446}{{\ttfamily 0808.1446}}].

\bibitem{Cohen:2010mi}
T.~Cohen, H.~Elvang and M.~Kiermaier, \emph{{On-shell constructibility of tree amplitudes in general field theories}}, \href{https://doi.org/10.1007/JHEP04(2011)053}{\emph{JHEP} {\bfseries 04} (2011) 053} [\href{https://arxiv.org/abs/1010.0257}{{\ttfamily 1010.0257}}].

\bibitem{Elvang:2013cua}
H.~Elvang and Y.-t.~Huang, \emph{{Scattering Amplitudes}},  \href{https://arxiv.org/abs/1308.1697}{{\ttfamily 1308.1697}}.

\bibitem{Henn:2014yza}
J.M.~Henn and J.C.~Plefka, \emph{{Scattering Amplitudes in Gauge Theories}}, vol.~883, Springer, Berlin (2014), \href{https://doi.org/10.1007/978-3-642-54022-6}{10.1007/978-3-642-54022-6}.

\bibitem{Cheung:2015ota}
C.~Cheung, K.~Kampf, J.~Novotny, C.-H.~Shen and J.~Trnka, \emph{{On-Shell Recursion Relations for Effective Field Theories}}, \href{https://doi.org/10.1103/PhysRevLett.116.041601}{\emph{Phys. Rev. Lett.} {\bfseries 116} (2016) 041601} [\href{https://arxiv.org/abs/1509.03309}{{\ttfamily 1509.03309}}].

\bibitem{Cheung:2017pzi}
C.~Cheung, \emph{{TASI Lectures on Scattering Amplitudes}},  in \emph{{Proceedings, Theoretical Advanced Study Institute in Elementary Particle Physics : Anticipating the Next Discoveries in Particle Physics (TASI 2016)}: {Boulder, CO, USA, June 6-July 1, 2016}}, R.~Essig and I.~Low, eds., pp.~571--623 (2018), \href{https://doi.org/10.1142/9789813233348_0008}{DOI} [\href{https://arxiv.org/abs/1708.03872}{{\ttfamily 1708.03872}}].

\bibitem{Arkani-Hamed:2017jhn}
N.~Arkani-Hamed, T.-C.~Huang and Y.-t.~Huang, \emph{{Scattering amplitudes for all masses and spins}}, \href{https://doi.org/10.1007/JHEP11(2021)070}{\emph{JHEP} {\bfseries 11} (2021) 070} [\href{https://arxiv.org/abs/1709.04891}{{\ttfamily 1709.04891}}].

\bibitem{Franken:2019wqr}
R.~Franken and C.~Schwinn, \emph{{On-shell constructibility of Born amplitudes in spontaneously broken gauge theories}}, \href{https://doi.org/10.1007/JHEP02(2020)073}{\emph{JHEP} {\bfseries 02} (2020) 073} [\href{https://arxiv.org/abs/1910.13407}{{\ttfamily 1910.13407}}].

\bibitem{Ballav:2020ese}
S.~Ballav and A.~Manna, \emph{{Recursion relations for scattering amplitudes with massive particles}}, \href{https://doi.org/10.1007/JHEP03(2021)295}{\emph{JHEP} {\bfseries 03} (2021) 295} [\href{https://arxiv.org/abs/2010.14139}{{\ttfamily 2010.14139}}].

\bibitem{Wu:2021nmq}
C.~Wu and S.-H.~Zhu, \emph{{Massive on-shell recursion relations for n-point amplitudes}}, \href{https://doi.org/10.1007/JHEP06(2022)117}{\emph{JHEP} {\bfseries 06} (2022) 117} [\href{https://arxiv.org/abs/2112.12312}{{\ttfamily 2112.12312}}].

\bibitem{Travaglini:2022uwo}
G.~Travaglini et~al., \emph{{The SAGEX review on scattering amplitudes}}, \href{https://doi.org/10.1088/1751-8121/ac8380}{\emph{J. Phys. A} {\bfseries 55} (2022) 443001} [\href{https://arxiv.org/abs/2203.13011}{{\ttfamily 2203.13011}}].

\bibitem{Ema:2024vww}
Y.~Ema, T.~Gao, W.~Ke, Z.~Liu, K.-F.~Lyu and I.~Mahbub, \emph{{Momentum shift and on-shell constructible massive amplitudes}}, \href{https://doi.org/10.1103/PhysRevD.110.105003}{\emph{Phys. Rev. D} {\bfseries 110} (2024) 105003} [\href{https://arxiv.org/abs/2403.15538}{{\ttfamily 2403.15538}}].

\bibitem{Ema:2024rss}
Y.~Ema, T.~Gao, W.~Ke, Z.~Liu, K.-F.~Lyu and I.~Mahbub, \emph{{Momentum shift and on-shell recursion relation for electroweak theory}}, \href{https://doi.org/10.1103/PhysRevD.110.105002}{\emph{Phys. Rev. D} {\bfseries 110} (2024) 105002} [\href{https://arxiv.org/abs/2407.14587}{{\ttfamily 2407.14587}}].

\bibitem{Gherghetta:2024tob}
T.~Gherghetta and W.~Ke, \emph{{Effective interactions and on-shell recursion relation for massive spin 3/2}}, \href{https://doi.org/10.1007/JHEP04(2025)014}{\emph{JHEP} {\bfseries 04} (2025) 014} [\href{https://arxiv.org/abs/2408.16065}{{\ttfamily 2408.16065}}].

\bibitem{Ema:2025qgd}
Y.~Ema, T.~Gao, W.~Ke, Z.~Liu and I.~Mahbub, \emph{{On-shell recursion relations for higher-spin Compton amplitudes}}, \href{https://doi.org/10.1007/JHEP01(2026)069}{\emph{JHEP} {\bfseries 01} (2026) 069} [\href{https://arxiv.org/abs/2506.02106}{{\ttfamily 2506.02106}}].

\bibitem{Trott:2026ozo}
T.~Trott, \emph{{Supersymmetry, Supergravity and the Consistency of On-Shell Massive Superamplitudes}},  \href{https://arxiv.org/abs/2608.04080}{{\ttfamily 2608.04080}}.

\bibitem{Jeong:2026dzt}
J.~Jeong, P.~Ko and Y.-H.~Zheng, \emph{{Explicit Conditions for Diagnosing Tree-Level Unitarity}},  \href{https://arxiv.org/abs/2605.12057}{{\ttfamily 2605.12057}}.

\bibitem{Bachu:2019ehv}
B.~Bachu and A.~Yelleshpur, \emph{{On-Shell Electroweak Sector and the Higgs Mechanism}}, \href{https://doi.org/10.1007/JHEP08(2020)039}{\emph{JHEP} {\bfseries 08} (2020) 039} [\href{https://arxiv.org/abs/1912.04334}{{\ttfamily 1912.04334}}].

\bibitem{Liu:2022alx}
D.~Liu and Z.~Yin, \emph{{Gauge invariance from on-shell massive amplitudes and tree-level unitarity}}, \href{https://doi.org/10.1103/PhysRevD.106.076003}{\emph{Phys. Rev. D} {\bfseries 106} (2022) 076003} [\href{https://arxiv.org/abs/2204.13119}{{\ttfamily 2204.13119}}].

\bibitem{Christensen:2024bdt}
N.~Christensen, \emph{{A Complete Set of 4-Point Amplitudes in the Constructive Standard Model}},  \href{https://arxiv.org/abs/2403.07978}{{\ttfamily 2403.07978}}.

\bibitem{Christensen:2024xzs}
N.~Christensen, \emph{{Perturbative Unitarity and the 4-Point Vertices in the Constructive Standard Model}},  \href{https://arxiv.org/abs/2403.07977}{{\ttfamily 2403.07977}}.

\bibitem{Christensen:2024nre}
N.~Christensen, \emph{{Field-theory action for the constructive standard model}}, \href{https://doi.org/10.1103/PhysRevD.110.105008}{\emph{Phys. Rev. D} {\bfseries 110} (2024) 105008} [\href{https://arxiv.org/abs/2408.13303}{{\ttfamily 2408.13303}}].

\bibitem{Ni:2024yrr}
Y.-H.~Ni, Y.-N.~Wang, C.~Wu and J.-H.~Yu, \emph{{Extended Poincare Symmetry Dictates Massive Scattering Amplitudes}},  \href{https://arxiv.org/abs/2412.03762}{{\ttfamily 2412.03762}}.

\bibitem{Ni:2025xkg}
Y.-H.~Ni, Y.-N.~Wang, C.~Wu and J.-H.~Yu, \emph{{Massive Helicity-Chirality Spinor Formalism from Massless Amplitudes with On-shell Mass Insertion}},  \href{https://arxiv.org/abs/2501.09062}{{\ttfamily 2501.09062}}.

\bibitem{Croft:2026eio}
M.~Croft and N.~Christensen, \emph{{The Scattering Algebra of Physical Space: Squared Massive Constructive Amplitudes}}, \href{https://doi.org/10.1007/s00006-025-01435-1}{\emph{Adv. Appl. Clifford Algebras} {\bfseries 36} (2026) 17} [\href{https://arxiv.org/abs/2601.16814}{{\ttfamily 2601.16814}}].

\bibitem{Ni:2026wiz}
Y.-H.~Ni, C.~Wu and J.-H.~Yu, \emph{{Massless-Massive Amplitude Correspondence I: Helicity-chirality Matching and On-shell Higgsing}},  \href{https://arxiv.org/abs/2601.10620}{{\ttfamily 2601.10620}}.

\bibitem{Ni:2026mia}
Y.-H.~Ni, C.~Wu and J.-H.~Yu, \emph{{Massless-Massive Amplitude Correspondence II: Constructive Massive Amplitudes in Standard Model}},  \href{https://arxiv.org/abs/2601.10622}{{\ttfamily 2601.10622}}.

\bibitem{Trott:2026cjj}
T.~Trott, \emph{{Consistent Scattering Amplitudes, Yang-Mills, the Higgs Mechanism and the EFTs Beyond}},  \href{https://arxiv.org/abs/2605.04152}{{\ttfamily 2605.04152}}.

\bibitem{Wang:2026qhv}
Y.-N.~Wang, C.~Wu and J.-H.~Yu, \emph{{Massive On-shell Splitting Functions in Spinor-Helicity Formalism}},  \href{https://arxiv.org/abs/2607.15191}{{\ttfamily 2607.15191}}.

\bibitem{LlewellynSmith:1973yud}
C.H.~Llewellyn~Smith, \emph{{High-Energy Behavior and Gauge Symmetry}}, \href{https://doi.org/10.1016/0370-2693(73)90692-8}{\emph{Phys. Lett. B} {\bfseries 46} (1973) 233}.

\bibitem{Cornwall:1974km}
J.M.~Cornwall, D.N.~Levin and G.~Tiktopoulos, \emph{{Derivation of Gauge Invariance from High-Energy Unitarity Bounds on the s Matrix}}, \href{https://doi.org/10.1103/PhysRevD.10.1145}{\emph{Phys. Rev. D} {\bfseries 10} (1974) 1145}.

\bibitem{Lee:1977eg}
B.W.~Lee, C.~Quigg and H.B.~Thacker, \emph{{Weak Interactions at Very High-Energies: The Role of the Higgs Boson Mass}}, \href{https://doi.org/10.1103/PhysRevD.16.1519}{\emph{Phys. Rev. D} {\bfseries 16} (1977) 1519}.

\bibitem{Chanowitz:1985hj}
M.S.~Chanowitz and M.K.~Gaillard, \emph{{The TeV Physics of Strongly Interacting W's and Z's}}, \href{https://doi.org/10.1016/0550-3213(85)90580-2}{\emph{Nucl. Phys. B} {\bfseries 261} (1985) 379}.

\bibitem{Gounaris:1986cr}
G.J.~Gounaris, R.~Kogerler and H.~Neufeld, \emph{{Relationship Between Longitudinally Polarized Vector Bosons and their Unphysical Scalar Partners}}, \href{https://doi.org/10.1103/PhysRevD.34.3257}{\emph{Phys. Rev. D} {\bfseries 34} (1986) 3257}.

\bibitem{Bagger:1989fc}
J.~Bagger and C.~Schmidt, \emph{{Equivalence Theorem Redux}}, \href{https://doi.org/10.1103/PhysRevD.41.264}{\emph{Phys. Rev. D} {\bfseries 41} (1990) 264}.

\bibitem{He:1992nga}
H.-J.~He, Y.-P.~Kuang and X.-y.~Li, \emph{{On the precise formulation of equivalence theorem}}, \href{https://doi.org/10.1103/PhysRevLett.69.2619}{\emph{Phys. Rev. Lett.} {\bfseries 69} (1992) 2619}.

\bibitem{Nardi:2024tce}
F.~Nardi, L.~Ricci and A.~Wulzer, \emph{{Low-virtuality splitting in the Standard Model}},  \href{https://arxiv.org/abs/2405.08220}{{\ttfamily 2405.08220}}.

\bibitem{Dittmaier:2025htf}
S.~Dittmaier and M.~Reyer, \emph{{Electroweak splitting functions in the Standard Model and beyond}}, \href{https://doi.org/10.1007/JHEP01(2026)119}{\emph{JHEP} {\bfseries 01} (2026) 119} [\href{https://arxiv.org/abs/2507.06568}{{\ttfamily 2507.06568}}].

\bibitem{Wulzer:2013mza}
A.~Wulzer, \emph{{An Equivalent Gauge and the Equivalence Theorem}}, \href{https://doi.org/10.1016/j.nuclphysb.2014.05.021}{\emph{Nucl. Phys. B} {\bfseries 885} (2014) 97} [\href{https://arxiv.org/abs/1309.6055}{{\ttfamily 1309.6055}}].

\bibitem{Chen:2016wkt}
J.~Chen, T.~Han and B.~Tweedie, \emph{{Electroweak Splitting Functions and High Energy Showering}}, \href{https://doi.org/10.1007/JHEP11(2017)093}{\emph{JHEP} {\bfseries 11} (2017) 093} [\href{https://arxiv.org/abs/1611.00788}{{\ttfamily 1611.00788}}].

\bibitem{Chen:2019dkx}
J.~Chen, \emph{{On the Feynman Rules of Massive Gauge Theory in Physical Gauges}},  \href{https://arxiv.org/abs/1902.06738}{{\ttfamily 1902.06738}}.

\bibitem{Cuomo:2019siu}
G.~Cuomo, L.~Vecchi and A.~Wulzer, \emph{{Goldstone Equivalence and High Energy Electroweak Physics}}, \href{https://doi.org/10.21468/SciPostPhys.8.5.078}{\emph{SciPost Phys.} {\bfseries 8} (2020) 078} [\href{https://arxiv.org/abs/1911.12366}{{\ttfamily 1911.12366}}].

\bibitem{Chen:2022gxv}
J.~Chen, K.~Hagiwara, J.~Kanzaki and K.~Mawatari, \emph{{Helicity amplitudes without gauge cancellation for electroweak processes}}, \href{https://doi.org/10.1140/epjc/s10052-023-12093-7}{\emph{Eur. Phys. J. C} {\bfseries 83} (2023) 922} [\href{https://arxiv.org/abs/2203.10440}{{\ttfamily 2203.10440}}].

\bibitem{Benincasa:2007xk}
P.~Benincasa and F.~Cachazo, \emph{{Consistency Conditions on the S-Matrix of Massless Particles}},  \href{https://arxiv.org/abs/0705.4305}{{\ttfamily 0705.4305}}.

\bibitem{McGady:2013sga}
D.A.~McGady and L.~Rodina, \emph{{Higher-spin massless S-matrices in four-dimensions}}, \href{https://doi.org/10.1103/PhysRevD.90.084048}{\emph{Phys. Rev. D} {\bfseries 90} (2014) 084048} [\href{https://arxiv.org/abs/1311.2938}{{\ttfamily 1311.2938}}].

\end{thebibliography}\endgroup

\end{document}